\documentclass[preprint]{aastex63}

\usepackage{amsmath}
\usepackage{amsfonts}

\begin{document}

\title{Resonant Chains and the Convergent Migration of Planets in Protoplanetary Disks}
\author{Ka Ho Wong}
\affiliation{Department of Earth Sciences, The University of Hong Kong,
  Pokfulam Road, Hong Kong}
\author[0000-0003-1930-5683]{Man Hoi Lee}
\affiliation{Department of Earth Sciences, The University of Hong Kong,
  Pokfulam Road, Hong Kong}
\affiliation{Department of Physics, The University of Hong Kong,
  Pokfulam Road, Hong Kong}

\begin{abstract}
An increasing number of compact planetary systems with multiple
planets in a resonant chain have been detected.
The resonant chain must be maintained by convergent migration of the
planets due to planet-disk interactions if it is formed before the
dispersal of the protoplanetary gas disk.
For type I migration in an adiabatic disk, we show that an analytic
criterion for convergent migration can be developed by requiring that
any part of the resonant chain should be convergently migrating toward
the remaining part.
The criterion depends primarily on the logarithmic gradients $\alpha$
and $\beta$ of the surface density and temperature profiles of the
disk, respectively, and it is independent of the absolute values of
the surface density and temperature.
The analytic criterion is applied to the Kepler-60, Kepler-80,
Kepler-223, TOI-178, and TRAPPIST-1 systems.
Due to the variation of planetary masses within the resonant chains,
we find that convergent migration typically requires rather extreme
values of $(\alpha, \beta)$ that have little or no overlap with common
disk models.
Finally, we show that there is an empirical relationship between the
distance of the innermost planet from the central star and the stellar
mass for the observed resonant chain systems, which supports the idea
that the resonant chains are formed and maintained by stalling the
migration of the innermost planet near the inner edge of the disk
truncated by the magnetic fields of the protostar.
\end{abstract}

\section{Introduction}
\label{sec:Introduction}

The first mean-motion resonance (MMR) in an extrasolar planetary system
was discovered around the star GJ~876, with planets b and c in 2:1
resonance \citep{Marcy2001}.
After two additional planets were discovered in this system, it also
became the first system with a three-body resonance, with planets b,
c, and e in a 4:2:1 Laplace resonance \citep{Rivera2010, Batygin2015b,
  Millholland2018}.
Many other systems with three or more planets in resonant chains have
since been discovered, particularly in transit surveys such as Kepler
and TESS.
Some of the famous resonant chain systems are Kepler-60
\citep{Steffen2013}, Kepler-80 \citep{MacDonald2016, Shallue2018},
Kepler-223 \citep{Mills2016}, and TRAPPIST-1 \citep{Gillon2017, Luger2017}.

Convergent migration is required for MMR capture \citep{Goldreich1965,
  Murray1999}.
It is believed that planet-disk interaction is the dominant migration
mechanism for assembling MMR and resonant chains of planets.
\cite{Lee2002} have shown that disk-induced migration is able to put
GJ~876 b and c into the observed MMR with suitable eccentricity
damping.
For some of the observed resonant chains, migration simulations have
also successfully assembled multiple planets into the observed
resonant chains (e.g., \citealt{Tamayo2017, Delisle2017,
  MacDonald2018}).
However, as pointed out by \cite{Papaloizou2018}, most of these
simulations applied inward migration on the outermost planet only.
Such a prescription ensures convergent migration and almost always
results in a resonant chain system if the migration is sufficiently
slow.
The inner planets will be captured into MMR, and the resonant chain
will be assembled from the outermost to the innermost planet.
However, this may not reflect the real situation where all planets are
embedded in the disk and are migrating individually.

Disk-induced planetary migration can be classified primarily into type
I and type II, depending on whether the planetary body is massive
enough to open a gap in the co-orbital region \citep{Kley2012}.
Type I migration is valid for the Earth-mass planets involved in many
observed resonant chains.
Since type I torque depends on planetary mass and various disk
properties (e.g., \citealt{Paardekooper2010}), we should examine what
kind of protoplanetary disk would allow convergent migration.
\cite{Batygin2015} has derived a criterion for convergent type I
migration of two planets in a classical Mestel disk,
\begin{equation} \label{eqs:Bat_crit}
m_{2}>\sqrt{\frac{a_2}{a_1}} m_1 ,
\end{equation}
where $m_1$ and $m_2$ are the masses of the inner and outer planets,
respectively, and $a_1$ and $a_2$ refer to their orbital semimajor
axes.\footnote{
\cite{Batygin2015} wrote that the necessary requirement for convergent
migration is $m_2 > \zeta m_1$, where $\zeta = \sqrt{a_1/a_2}$.
It is, however, clear from his Equation (47) for the type I migration
timescale that the necessary requirement should be $m_2 > \zeta^{-1}
m_1$, which is our Equation (\ref{eqs:Bat_crit}).}
This criterion assumes (1) a disk surface density profile $\Sigma =
\Sigma_0 a_0/a$, where $\Sigma_0$ is the surface density at the
reference semimajor axis $a_0$; (2) a constant disk aspect ratio
$h = H/a$ throughout the disk; and (3) an isothermal equation of
state.
These assumptions make the criterion appear independent of the disk
parameters.
Equation (\ref{eqs:Bat_crit}) says that the outer planet has to be
more massive than the inner planet for convergent inward migration and
the subsequent MMR assembly to happen.
Such criterion on the planetary masses is not satisfied in many
observed resonant chain systems.
Take TRAPPIST-1 as an example: the planet h has a mass of $0.3
M_{\oplus}$, which is significantly less massive than the inner planet
g of $1.34 M_{\oplus}$ \citep{Grimm2018}.
Similar examples are seen in other resonant chain systems.
This implies that these resonant chains are unlikely to be assembled in a
classical Mestel disk.

Regardless of how a resonant chain is assembled, if it is formed in
the gas disk phase and coexists with the disk, the resonant chain
must be maintained during migration, which requires convergent
migration within the chain.
In this study, we consider type I migration in an adiabatic disk and
investigate what disk properties are required to maintain a resonant
chain by convergent migration.
The adopted disk and migration models are described in Section
\ref{sec:Disk Model}.
In Section \ref{sec:Two Planets}, we analytically derive the necessary
conditions for the convergent migration of a pair of planets, which
are shown to be accurate by comparison with numerical results.
The criterion is generalized to the migration of multiple planets in a
resonant chain in Section \ref{sec:Resonant Chains}.
The analytic criterion is tested numerically for the four-planet
resonant chain system Kepler-223 and then applied to other observed
resonant chains.
We find that rather extreme disk parameters, far from the usual disk
models, are required for convergent migration.
The effects of uncertainties in planetary masses are discussed in
Section \ref{sec:Uncertainties}.
\cite{Terquem2007}, \cite{Cossou2013}, \cite{Brasser2018}, and
\cite{Huang2022} have shown that an inner disk edge can stall
migration and facilitate the assembly of an MMR pair and resonant chain
(see also \citealt{Lithesis2014} and \citealt{Batygin2020} for a
similar scenario for the Laplace resonance of the Galilean satellites
of Jupiter).
In Section \ref{sec:Scaling}, we find a simple relationship between
the mass of the host star and the orbital semimajor axis of the
innermost planet in the observed resonant chains, which supports
this idea.
Our conclusions are summarized in Section \ref{sec:Conclusions}.

\section{Disk and Migration Models}
\label{sec:Disk Model}

We adopt an adiabatic disk model with surface density profile $\Sigma
= \Sigma_0 (a_0/a)^\alpha$ and temperature profile $T = T_0
(a_0/a)^\beta$.
Since the scale height $H = c_{s}/n$, where the sound speed $c_{s}
\propto T^{1/2}$ and the mean motion $n \propto a^{-3/2}$, the reduced
scale height $h = H/a = h_0 (a/a_0)^{(-\beta+1)/2}$.
Thus, the disk has five parameters $(\alpha, \beta, \gamma, \Sigma_0,
h_0)$.
Hereafter, we assume that the adiabatic index $\gamma = 5/3$.
One can also think of $\alpha$ and $\beta$ as the local power-law
indices of the surface density and temperature profiles:  $\alpha=-d
\ln \Sigma/d \ln a$ and $\beta = -d \ln T /d \ln a$.

\cite{Paardekooper2010} have derived a formula for the unsaturated
nonlinear type I migration torque from an adiabatic disk on the
planet:
\begin{eqnarray}
\Gamma &=& \frac{1}{\gamma} \left[-2.5 - 1.7 \beta + 0.1\alpha +
           1.1\left(\frac{3}{2}-\alpha\right) + 7.9
           \frac{\beta-(\gamma-1)\alpha}{\gamma}\right] \Gamma_0 ,
\label{eqs:Paar_torque_1a} \\
       &=& \frac{1}{\gamma} \left[-0.85 - \left(7.9
           \frac{\gamma-1}{\gamma}+1\right)\alpha +
           \left(\frac{7.9}{\gamma}-1.7\right)\beta\right] \Gamma_0 ,
\label{eqs:Paar_torque_1b}
\end{eqnarray}
where
\begin{equation} \label{eqs:Paar_torque_2}
\Gamma_0 = \left( \frac{m}{m_{0}}\right)^2 h^{-2} \Sigma a^4 n^2,
\end{equation}
and $m$ and $m_0$ refer to the masses of the planet and the star,
respectively.
There are three components to the torque in Equation
(\ref{eqs:Paar_torque_1a}): (1) the first term with the coefficient
$-2.5 - 1.7 \beta + 0.1\alpha$ is the Lindblad torque, (2) the second
term with the coefficient $1.1(3/2 - \alpha)$ is the barotropic term
of the nonlinear horseshoe torque, and (3) the last term is the
entropy-related term of the nonlinear horseshoe torque.

\cite{Paardekooper2011} have studied the effects of viscous and
thermal diffusion on the nonlinear corotation torque in Equations
(\ref{eqs:Paar_torque_1a})--(\ref{eqs:Paar_torque_2}).
Equations (\ref{eqs:Paar_torque_1a})--(\ref{eqs:Paar_torque_2}) are valid
if the viscous saturation parameter $p_\nu$ and the thermal
saturation parameter $p_\chi$ are $\sim 0.5$.
In the limit $p_\nu$ and $p_\chi \ll 1$, the nonlinear horseshoe
torque is replaced by the linear corotation torque.
In the limit $p_\nu$ and $p_\chi \gg 1$, the corotation torque
saturates, and only the Lindblad torque remains.
The viscous saturation parameter $p_\nu = 2 \sqrt{k x_s^3}/3$, where
$k = a^2 n/(2 \pi \nu)$, $x_s = 1.1 \gamma^{-1/4}
(m/m_0)^{1/2} h^{-1/2}$ is the dimensionless half-width of the horseshoe
region, and $\nu$ is the viscosity.
For $\alpha_\nu$ viscosity with $\nu = \alpha_\nu c_s H$
\citep{Shakura1973}, $p_\nu \approx 0.25 (m/m_0)^{3/4} \alpha_\nu^{-1/2} h^{-7/4}$.
For the planets in the resonant chain systems that we consider below,
$m/m_0 = 1.0$--$4.6 \times 10^{-5}$ and $p_\nu = 0.27$--$0.85
(\alpha_\nu/0.001)^{-1/2} (h/0.05)^{-7/4}$.
The thermal saturation parameter $p_\chi = 3 p_\nu \sqrt{\nu/\chi}/2 =
\sqrt{a^2 n x_s^3/(2 \pi\chi)}$, where $\chi = 16 \gamma (\gamma -
1)\sigma_{\text{SB}} T^4/(3 \kappa \Sigma^2 n^2)$ is the thermal
diffusion coefficient,  $\sigma_{\text{SB}}$ is the Stefan-Boltzmann
constant, and $\kappa$ is the opacity.
The value of $p_\chi$ depends on the actual values of $a$, $n$,
$\Sigma$, and $\kappa$, not just the dimensionless parameters $m/m_0$
and $h$.
If we assume that $\Sigma_0 = 1700\,\text{g}\,\text{cm}^{-2}$ and $h_0
= 0.05$ at $a_0 = 1\,$au, similar to the minimum mass solar nebula
(MMSN), and that $\kappa = 1\,\text{cm}^2\,\text{g}^{-1}$, $p_\chi =
0.21$--$0.65$ for the planets under consideration.
Based on these estimates of $p_\nu$ and $p_\chi$, the unsaturated
nonlinear type I migration torque should be applicable to the planets
considered in this paper.

The forced migration rate from the tidal torque on the planet is
\begin{equation} \label{eqs:adot1}
(\dot{a}/a)_{\text{f}} = 2 \Gamma/(m \sqrt{G m_0 a}) ,
\end{equation}
where the subscript f indicates forced migration.
The correction from eccentricity \citep{Goldreich2014} is neglected to
simplify derivation later, and it has little effect on the discussion
in this paper on the convergent properties of migration.

\begin{figure}
  \centering
  \gridline{\fig{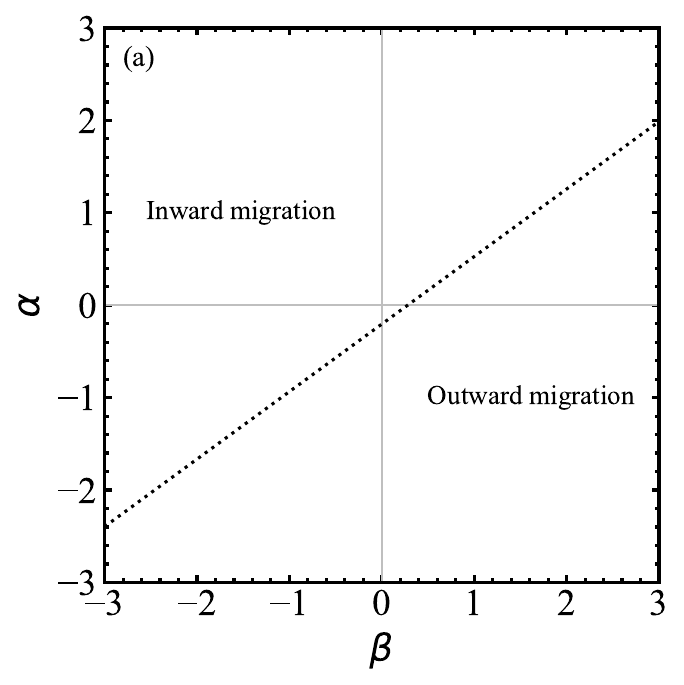}{0.4\textwidth}{}
            \fig{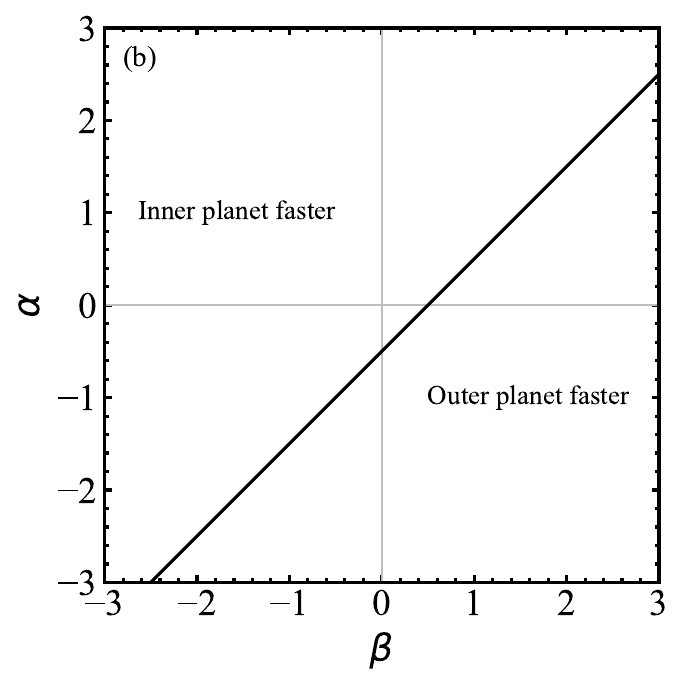}{0.4\textwidth}{}
    }
  \gridline{\fig{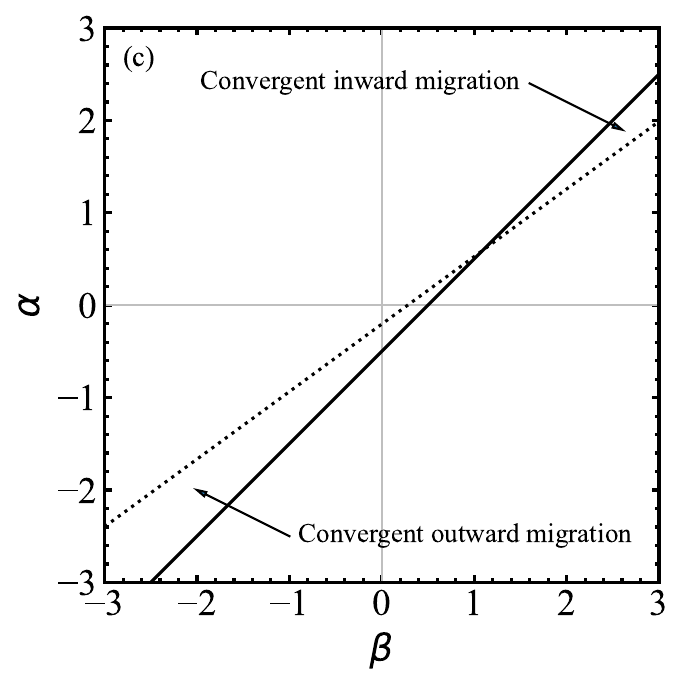}{0.4\textwidth}{}}
\caption{(a) Boundary between inward and outward migration according
to Equation (\ref{eqs:disk_constraint}) for an adiabatic disk with
power-law indices $\alpha$ for surface density and $\beta$ for
temperature.
(b) Boundary between faster forced migration of inner planet and
faster forced migration of outer planet according to Equations
(\ref{eqs:crit_m}) and (\ref{eqs:crit_m2}) for $m_1 = m_2$.
(c) Combined plot, with the convergent outward and inward migration
regions corresponding to the regions bounded by the red and blue lines
in Figure \ref{fig:2_pl_sus}.}
\label{fig:2_pl_analy}
\end{figure}

If the expression in the square brackets in Equation
(\ref{eqs:Paar_torque_1b}) is positive (negative), there is a positive
(negative) torque and outward (inward) migration.
The boundary between inward and outward migration is simply deduced by
setting $\Gamma$ in Equation (\ref{eqs:Paar_torque_1b}) to zero:
\begin{equation}
  \alpha = \left[-0.85 + (7.9/\gamma - 1.7)\beta\right]/\left[7.9
    (\gamma - 1)/\gamma + 1\right] .
\label{eqs:disk_constraint}
\end{equation}
This line is shown in Figure \ref{fig:2_pl_analy}(a) in the $(\alpha,
\beta)$ space with $\gamma = 5/3$.
Inward (outward) migration occurs in the region above (below) the
line.
Note that the condition for inward/outward migration is independent
of the mass $m$ and location $a$ of the planet.

\section{Conditions for Convergent Migration of Two Planets}
\label{sec:Two Planets}

For an inner planet of mass $m_1$ and an outer planet of mass $m_2$,
irrespective of whether the forced migration is inward or outward, the
dividing line between the situation where the forced migration of the
inner planet is faster than that of the outer planet and vice versa is
$\lvert \dot{a}_1/a_1 \rvert_{\text{f}} = \lvert \dot{a}_2/a_2 \rvert_{\text{f}}$.
According to Equations (\ref{eqs:Paar_torque_1a})--(\ref{eqs:adot1}),
the torque exerted on a planet of mass $m$ at $a$ is $\Gamma \propto
\Gamma_0 \propto m^2 h^{-2} \Sigma a^4 n^2 \propto m^2
a^{-\alpha + \beta}$, and the forced migration rate $({\dot
  a}/a)_{\text{f}} \propto m a^{-\alpha + \beta - 1/2}$.
Thus, the condition $\lvert \dot{a}_1/a_1 \rvert_{\text{f}} = \lvert
\dot{a}_2/a_2 \rvert_{\text{f}}$ can be written as
\begin{equation} \label{eqs:crit_m}
m_2 = \left( \frac{P_2}{P_1} \right) ^{(1+2\alpha-2\beta)/3} m_1 ,
\end{equation}
or
\begin{equation} \label{eqs:crit_m2}
\alpha - \beta =
  - \frac{1}{2} + \frac{3}{2} \frac{\ln(m_2/m_1)}{\ln(P_2/P_1)} ,
\end{equation}
where $P_1$ and $P_2$ are the orbital periods.
This line is shown in Figure \ref{fig:2_pl_analy}(b) in the
$(\alpha, \beta)$ space for an example of two equal-mass planets, and
it is independent of the period ratio because $\ln(m_2/m_1) = 0$.
The forced migration of the inner (outer) planet is faster in the
region above (below) the line.
Note that Equation (\ref{eqs:crit_m}) for the Mestel disk with
constant $h$ (i.e., $\alpha = \beta = 1$) agrees with Equation
(\ref{eqs:Bat_crit}).

Two planets undergo convergent migration if their migration rates
satisfy $\lvert \dot{a}_1/a_1 \rvert_{\text{f}} <
\lvert \dot{a}_2/a_2 \rvert_{\text{f}}$ for inward migration or
$\lvert \dot{a}_1/a_1 \rvert_{\text{f}} >
\lvert \dot{a}_2/a_2 \rvert_{\text{f}}$ for outward migration.
Thus, the convergent migration zone is bounded by two lines, the disk
constraint in Equation (\ref{eqs:disk_constraint}) and the planetary
mass constraint in Equation (\ref{eqs:crit_m}) or (\ref{eqs:crit_m2}).
This is shown in Figure \ref{fig:2_pl_analy}(c) for the example
of two equal-mass planets, with the inward and outward convergent
migration regions labeled.
We stress again that convergent migration is only a necessary
condition to maintain MMR.
The capture probability also depends on the eccentricities and the
migration and eccentricity damping rates
\citep{Mustill2011, Goldreich2014, Deck2015, Batygin2023b, Huang2023,
Kajtazi2023}.

\begin{figure}
  \centering
  \gridline{\fig{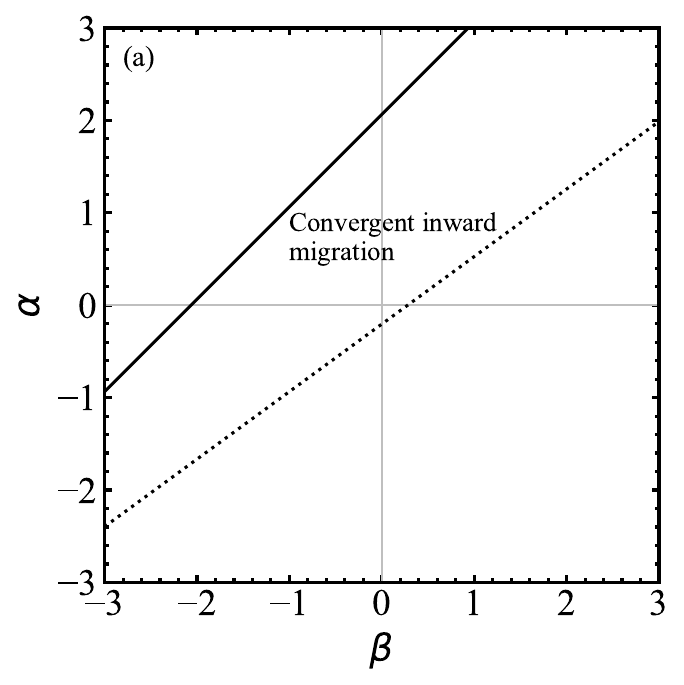}{0.4\textwidth}{}
            \fig{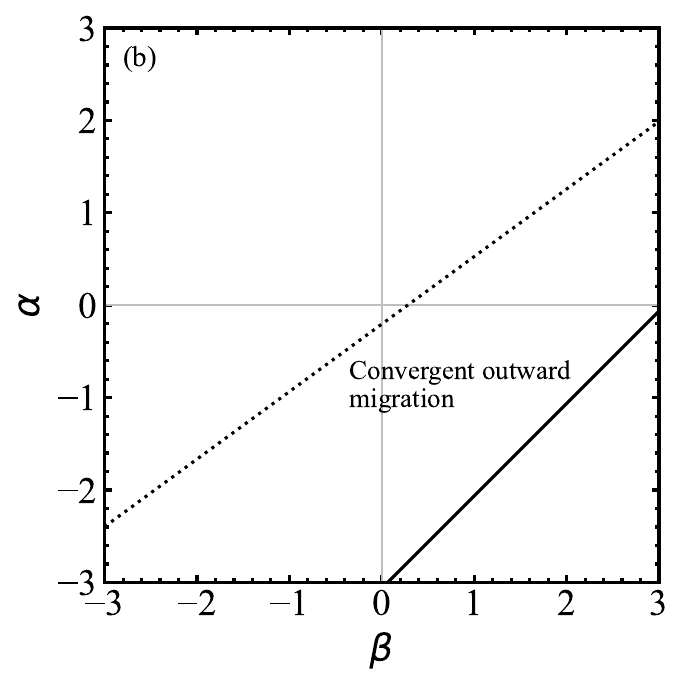}{0.4\textwidth}{}
    }
\caption{Same as Figure \ref{fig:2_pl_analy}(c) but for $P_2/P_1 =
3/2$ and (a) $m_2/m_1 = 2$ and (b) $m_2/m_1 = 1/2$. The convergent
inward and outward migration zones are marked.}
\label{fig:2_pl_mass_analy}
\end{figure}

A system with $m_2$ more massive than $m_1$ should favor convergent
inward migration, while a system with a more massive $m_1$ should
favor convergent outward migration.
This intuition is verified with Equation (\ref{eqs:crit_m2}) and
demonstrated in Figure \ref{fig:2_pl_mass_analy}.
In Figure \ref{fig:2_pl_mass_analy}, the dotted and solid lines
have the same meaning as those in Figure \ref{fig:2_pl_analy} but now
for $P_2/P_1 = 3/2$ and $m_2/m_1 = 2$ in panel (a) and $m_2/m_1 = 1/2$
in panel (b).
We see a large convergent inward migration region and no convergent
outward migration region within the range of $(\alpha, \beta)$ shown
in panel (a), and vice versa in panel (b).

We have tested the validity of the analytic conditions in Equations
(\ref{eqs:disk_constraint}),
(\ref{eqs:crit_m}), and (\ref{eqs:crit_m2}) using numerical simulations.
We start with a system with two planets already in 3:2 MMR, which is
obtained from a prior migration simulation.
For the masses, we adopt $1.125 M_\odot$ for the central star
(which is the stellar mass of Kepler-223; see below) and $1
M_\earth$ for both planets.
The initial orbital elements are listed in Table
\ref{table:2_pl_init}.
The planets are initially in 3:2 MMR with low eccentricities,
antialigned periapses, and both MMR angles in libration.
For the disk parameters, we assume that $\Sigma_0 =
1700\,\text{g}\,\text{cm}^{-2}$ and $h_0 = 0.05$ at $a_0 = 1\,$au,
similar to the MMSN.
The disk parameters $\alpha$ and $\beta$ are surveyed uniformly in a
$30 \times 30$ grid from $-3$ to $+3$.
The adopted masses, $\Sigma_0$, and $h_0$ only affect the absolute
migration rate but not the sign of the relative migration rate and the
locations of the convergent and divergent migration zones in $(\alpha,
\beta)$ space.
We use the SyMBA integrator \citep{Lee1998} with imposed migration
following Equations (\ref{eqs:Paar_torque_1a})--(\ref{eqs:adot1}).
An eccentricity damping of $\dot{e}/e=-100 \lvert \dot{a}/a \rvert$ is
applied on both planets, which avoids the instability caused by high
eccentricity during convergent migration.
For each combination of $\alpha$ and $\beta$, the system is integrated
for one migration timescale $a_2/{\dot a}_2$ of the outer planet.
At the end of the simulation, we check whether the system is still
in the 3:2 MMR by examining the libration of the resonant angles.

\begin{deluxetable}{ccccccc}
\tablewidth{0pt}
\tablecaption{Initial Conditions of Two Planets in 3:2 MMR for Disk-driven Migration Simulations
\label{table:2_pl_init}}
\tablehead{
\colhead{Planet} & \colhead{Mass ($M_{\oplus}$)} & \colhead{$a$ (au)} & \colhead{$e$} & \colhead{$i,\Omega$} & \colhead{$\omega$} & \colhead{$M$}
}
\startdata
$m_1$ & 1 & 0.9994 & 0.01113 & $0\degr$ &  18.09$\degr$ & 353.9$\degr$\\
$m_2$ & 1 & 1.3097 & 0.01188 & $0\degr$ & 197.60$\degr$ & 296.6$\degr$\\
\enddata
\tablecomments{The orbital elements are semimajor axis $a$,
eccentricity $e$, inclination $i$, longitude of the ascending node
$\Omega$, argument of periapse $\omega$, and mean anomaly $M$.
The stellar mass is 1.125 $M_{\odot}$.}
\end{deluxetable}

\begin{figure}
\centering
\includegraphics[width=0.8\linewidth]{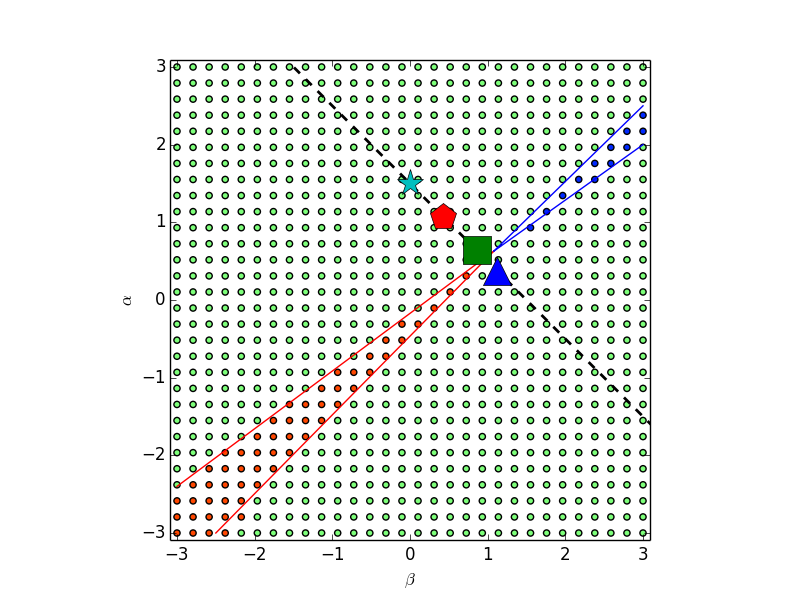}  
\caption{Numerical results in $(\alpha, \beta)$ space for the
migration of a pair of $1M_{\oplus}$ planets in  3:2 MMR.
Blue and red dots represent simulations where the MMR is maintained as
the planets move inward and outward, respectively.
Green dots represent the simulations where the MMR is broken.
The blue and red solid lines are the analytic estimation of the
convergent migration regions from Equations
(\ref{eqs:disk_constraint}),
(\ref{eqs:crit_m}), and (\ref{eqs:crit_m2}).
The black dashed line is $\alpha + \beta = 3/2$ for the steady-state,
constant $\alpha_\nu$-viscosity disk models, with the star for the
MMSN-like model and the other symbols for various regions of the
\cite{Garaud2007} disk model: pentagon for the weakly opaque region,
square for the strongly opaque region, and triangle for the viscously
heated region.}
\label{fig:2_pl_sus}
\end{figure}

The simulation results are shown in Figure \ref{fig:2_pl_sus}.
The green dots represent the systems where the 3:2 MMR is broken
during the simulation.
The red dots indicate the systems where the planets move outward and
the MMR is maintained.
The blue dots indicate the systems where the planets move inward and
the MMR is maintained.
The solid lines are the same lines shown in Figure
\ref{fig:2_pl_analy}(c) from  the analytic conditions in Equations
(\ref{eqs:disk_constraint}),
(\ref{eqs:crit_m}), and (\ref{eqs:crit_m2}).
The analytic theory and numerical simulations match quite well.
Even systems just outside the zones bounded by the two lines break
from 3:2 MMR, which means that a small relative divergent migration is
enough to break the MMR within one migration timescale.

For comparison, the dashed line in Figure \ref{fig:2_pl_sus}
represents the steady-state, constant $\alpha_\nu$-viscosity disk
models, which obey $\alpha + \beta = 3/2$ \citep{Shakura1973,
  Kretke2012}.
The dashed line crosses the outward convergent migration zone near the
intersection between inward and outward convergent migration zones,
and only a narrow range of such models with $\alpha \approx 0.5$ and
$\beta \approx 1.0$ have convergent migration that maintains MMR for a
pair of equal-mass planets.
Several specific constant $\alpha_\nu$-viscosity models are also
indicated: a disk with the same density profile as the MMSN
(star, $\alpha = 3/2$, $\beta = 0$), and the weakly opaque region
(pentagon, $\alpha = 15/14$, $\beta = 3/7$), strongly opaque
region (square, $\alpha = 9/14$, $\beta = 6/7$), and viscously
heated region (triangle, $\alpha = 3/8$, $\beta = 9/8$) of the
\cite{Garaud2007} disk model \citep{Kretke2012}.
None of these models are within the convergent migration zones for
equal-mass planets.
However, all but the viscously heated region would be within the
convergent inward migration zone for $m_2/m_1 = 2$ and 3:2 MMR (see
panel (a) of Figure \ref{fig:2_pl_mass_analy}).


\section{Generalization and Application to Resonant Chain Systems}
\label{sec:Resonant Chains}

In this section, we generalize the analytic criterion for convergent
migration of a pair of planets derived in the previous section to the
case of multiple planets in a resonant chain.
The Kepler-223 system is taken as an illustrative example to verify
the analytic criterion with numerical simulations.
We then apply the analytic criterion to the Kepler-60, Kepler-80,
TOI-178, and TRAPPIST-1 systems.
The results show that rather extreme values of $(\alpha, \beta)$ are
required to maintain the resonant chains in these systems.

\subsection{Convergent Migration Criterion for Multiple Planets in
  Resonant Chains}
\label{sec:Resonant Chain Criterion}

To construct the analytic criterion, one should distinguish between
the actual migration rate of a resonant chain and the forced
migration rate from the torque applied by the disk on each planet.
If a resonant chain is maintained by convergent migration within the
chain, the actual migration rate ${\dot a}/a$ must be the same for all
planets and the same as the comigration rate of the whole chain.
For a resonant chain of $N$ planets, the comigration rate
$({\dot a}/a)_{12...N}$ can be calculated by equating the work done by
the torques on the individual planets to the total orbital energy
change of the chain, with the assumption that the orbital period
ratios are fixed:
\begin{equation} \label{eqs:co_mig}
\left(\frac{{\dot a}}{a}\right)_{123\cdots N} =
\frac{\displaystyle
  \left(\frac{{\dot a}_1}{a_1}\right)_{\text{f}} +
  \left(\frac{m_2}{m_1}\right) \left(\frac{P_2}{P_1}\right)^{-2/3}
  \left(\frac{{\dot a}_2}{a_2}\right)_{\text{f}} +
  \cdots +
  \left(\frac{m_N}{m_1}\right) \left(\frac{P_N}{P_1}\right)^{-2/3}
  \left(\frac{{\dot a}_N}{a_N}\right)_{\text{f}}}
{\displaystyle
  1 +
  \left(\frac{m_2}{m_1}\right) \left(\frac{P_2}{P_1}\right)^{-2/3} +
  \cdots +
  \left(\frac{m_N}{m_1}\right) \left(\frac{P_N}{P_1}\right)^{-2/3}} ,
\end{equation}
where $({\dot a}_{i}/a_{i})_{\text{f}}$ is the forced migration rate
of the planet $i$ of mass $m_i$ and orbital period $P_i$.
Given a certain set of disk parameters $(\alpha, \beta, \gamma,
\Sigma_0, h_0)$, we can calculate the individual planet's forced
migration rate $({\dot a}_i/a_i)_{\text{f}}$ from Equations
(\ref{eqs:Paar_torque_1b})--(\ref{eqs:adot1}) and the comigration
rate $({\dot a}/a)_{12...N}$ from Equation (\ref{eqs:co_mig}) for any
planet pair, triplet, or the whole resonant chain in a disk, under the
assumption that they are locked in resonance.

It is straightforward to generalize the necessary criterion to
maintain a resonant pair to a resonant chain.
The idea is that any part of the resonant chain should be convergently
migrating toward the remaining part.
Taking a resonant chain of four planets as an example, the generalized
criterion for convergent inward migration is
\begin{equation} \label{eqs:co_mig_crit_in}
\lvert {\dot a}/a\rvert_{1} < \lvert {\dot a}/a\rvert_{234}
\quad \text{and} \quad
\lvert {\dot a}/a\rvert_{12} < \lvert {\dot a}/a\rvert_{34}
\quad \text{and} \quad
\lvert {\dot a}/a\rvert_{123} < \lvert {\dot a}/a\rvert_{4} .
\end{equation}
This means that planet 1 should be migrating at a rate slower than
that of the triplet 2-3-4.
Similarly, the migration rate of the planet pair 1-2 should be slower
than that of the planet pair 3-4, and the migration rate of the
triplet 1-2-3 should be slower than that of planet 4.
For outward migration, the inequalities are flipped, i.e.,
\begin{equation} \label{eqs:co_mig_crit_out}
\lvert {\dot a}/a\rvert_{1} > \lvert {\dot a}/a\rvert_{234}
\quad \text{and} \quad
\lvert {\dot a}/a\rvert_{12} > \lvert {\dot a}/a\rvert_{34}
\quad \text{and} \quad
\lvert {\dot a}/a\rvert_{123} > \lvert {\dot a}/a\rvert_{4} .
\end{equation}
The set of inequalities for inward or outward migration have to be
satisfied simultaneously because no part of the chain should be
divergently moving away from the rest of the chain.

A similar criterion for a planet pair leads to Equations
(\ref{eqs:crit_m}) and (\ref{eqs:crit_m2}) for the combination $\alpha -
\beta$ in the exponent.
For a resonant chain, since the actual migration rate of two or more
planets is given by the more complicated Equation (\ref{eqs:co_mig}),
it is not possible to reduce the criterion to a similarly simple form.
Each part of the resonant chain in Equations
(\ref{eqs:co_mig_crit_in}) and (\ref{eqs:co_mig_crit_out}) results in an
equation (e.g., $\lvert {\dot a}/a\rvert_{12} = \lvert
{\dot a}/a\rvert_{34}$) for $\alpha - \beta$, which can be solved
numerically and combined with the disk constraint in Equation
(\ref{eqs:disk_constraint}) to determine the inward and outward
convergent migration regions in the $(\alpha, \beta)$ space.
The constraints from all parts of the resonant chain can then be
combined to give the overlapping regions (if any) where the resonant
chain can be maintained by convergent migration.

\subsection{Application to Kepler-223} \label{sec:Application to Kepler-223}

\begin{deluxetable}{ccccccc}
\tablewidth{0pt}
\tablecaption{Initial Conditions of a Planetary System in 8:6:4:3
  Resonance for Disk-driven Migration Simulations
  \label{table:223_init}}
\tablehead{
  \colhead{Planet} & \colhead{Mass ($M_{\oplus}$)} & \colhead{$a$ (au)} & \colhead{$e$} & \colhead{$i,\Omega$} & \colhead{$\omega$} & \colhead{$M$}
  }
\startdata
b or 1 & 7.4 & 1.0274 & 0.01646 & $0\degr$ & 203.64$\degr$ &  76.18$\degr$\\
c or 2 & 5.1 & 1.2457 & 0.02493 & $0\degr$ &  27.78$\degr$ & 146.16$\degr$\\
d or 3 & 8.0 & 1.6337 & 0.00854 & $0\degr$ & 166.70$\degr$ &  73.27$\degr$\\
e or 4 & 4.8 & 1.9813 & 0.00967 & $0\degr$ & 321.51$\degr$ & 165.27$\degr$\\
\enddata
\tablecomments{Stellar mass is 1.125 $M_{\odot}$.}
\end{deluxetable}

\begin{figure}
  \centering
  \gridline{\fig{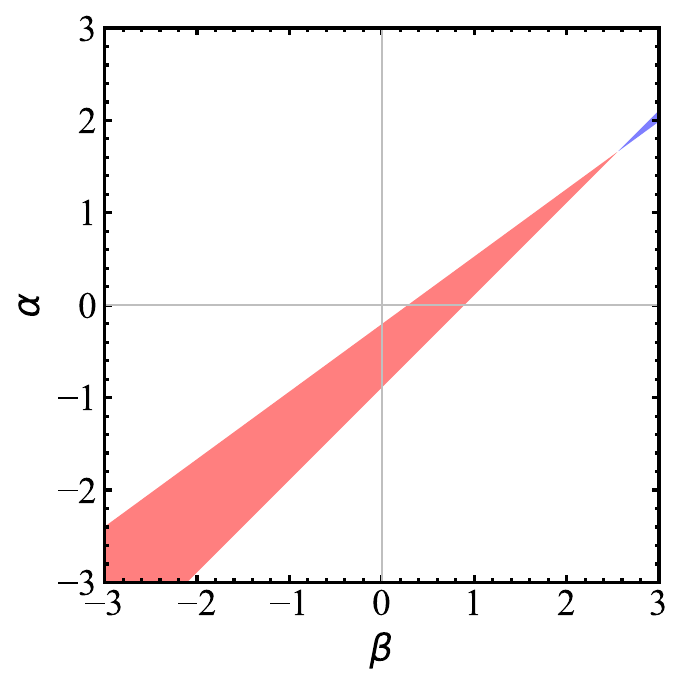}{0.4\textwidth}{(a) Planet b and triplet c-d-e}
          \fig{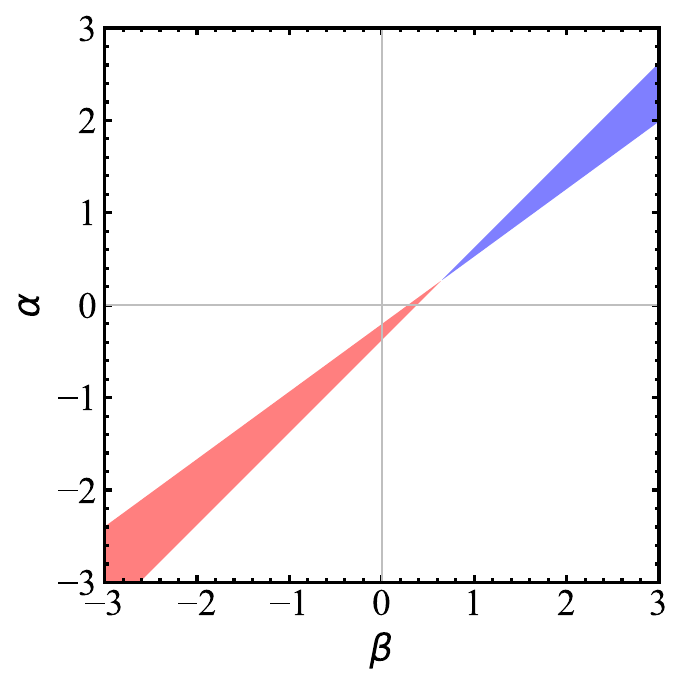}{0.4\textwidth}{(b) b-c and d-e pair}
  }
  \gridline{\fig{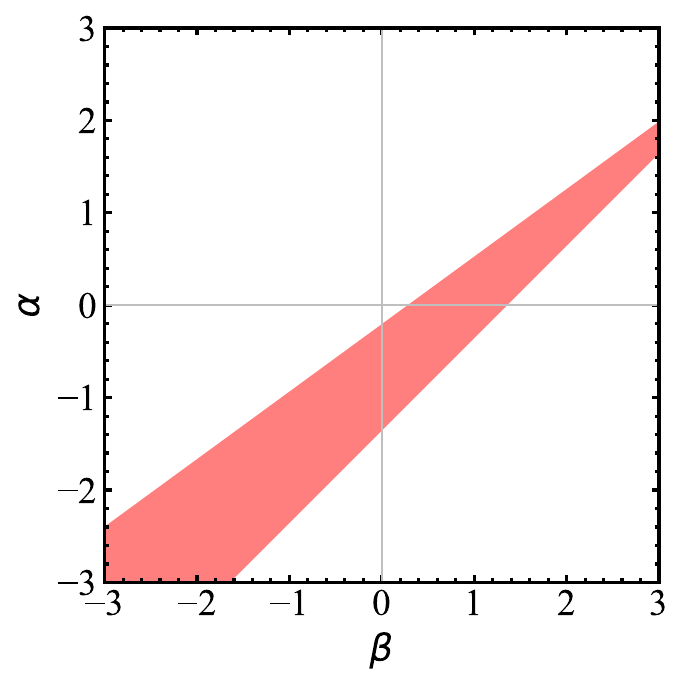}{0.4\textwidth}{(c) Triplet b-c-d and planet e}
          \fig{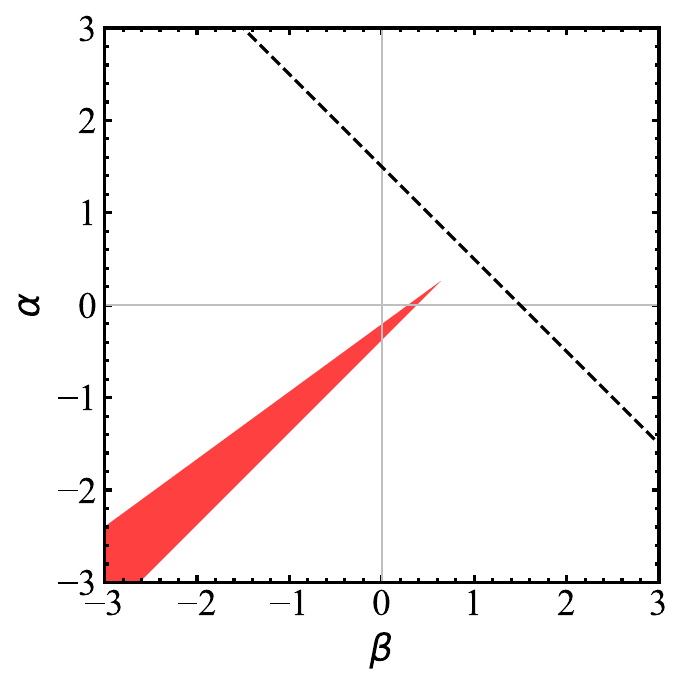}{0.4\textwidth}{(d) Combined result}
          }
\caption{Analytic results on Kepler-223 from Equations
  (\ref{eqs:disk_constraint}),
  (\ref{eqs:co_mig_crit_in}), and (\ref{eqs:co_mig_crit_out}), with panels
  (a), (b), and (c) showing the regions in $(\alpha, \beta)$ space
  with convergent migration between planet b and triplet c-d-e,
  between pair b-c and pair  d-e, and between triplet b-c-d and planet
  e, respectively, and panel (d) showing the combined result from the
  overlap of panels (a)--(c).
  Red indicates convergent outward migration, and blue
  indicates convergent inward migration.
  Panel (d) shows that the Kepler-223 resonant chain cannot be
  sustained by inward migration and that the convergent outward
  migration zone does not include the steady-state, constant
  $\alpha_\nu$-viscosity disk models (dashed line).
}
\label{fig:223_analy}
\end{figure}

Let us use the 8:6:4:3 resonant chain in Kepler-223 as an example.
We adopt a stellar mass of $1.125 M_{\odot}$ and the planetary
masses as listed in Table \ref{table:223_init} from \cite{Mills2016}.
Panels (a), (b), and (c) of Figure \ref{fig:223_analy} show the
convergent migration zones in the $(\alpha, \beta)$ space between
planet b and triplet c-d-e, between pair b-c and pair d-e, and between
triplet b-c-d and planet e, respectively.
The inward convergent zone is shaded in blue, and the outward
convergent zone is shaded in red.
In each panel, one can see that the convergent zones have the same
shapes as those for two planets in Figures \ref{fig:2_pl_analy} and
\ref{fig:2_pl_mass_analy}.
This is because one of the lines defining the convergent zones --- the
boundary between inward and outward migration in Equation
(\ref{eqs:disk_constraint}) --- is the same in all cases, while the
other line from a comparison of the migration rate of a part of the
resonant chain with the migration rate of the rest of the resonant
chain (Equations (\ref{eqs:co_mig_crit_in}) and
(\ref{eqs:co_mig_crit_out})) is simply shifted vertically (i.e., a
different value of $\alpha - \beta$) for each part of the resonant
chain.
In panel (c), one can see that there is no convergent inward migration
zone for triplet b-c-d and planet e in the region of $(\alpha, \beta)$
shown.
This is because planet e has a small mass of $4.8 M_{\oplus}$, which
makes it difficult for planet e to catch up with b-c-d in inward
migration, unless extreme values of $(\alpha, \beta)$ are adopted.
Panel (d) is the combined constraint, which shows the overlapping
region from panels (a), (b), and (c).
The analytic criterion indicates that the resonant chain in Kepler-223
cannot be maintained by inward convergent migration for any values of
$\alpha$ and $\beta$ between $-3$ and $+3$ and that it can be
maintained by outward convergent migration in a narrow region of
mostly negative values of $\alpha$ and $\beta$ (i.e., both surface
density and temperature increasing outward).

Numerical simulations with imposed migration have been carried out to
test the analytic criterion in Figure \ref{fig:223_analy}.
We start with a system with four planets assembled into the desired
8:6:4:3 resonant chain by a prior migration simulation.
The orbital parameters are listed in Table \ref{table:223_init}.
The three-body resonance or Laplace angles $\theta_{\text{L,in}} =
-\lambda_{1} + 2\lambda_{2} - \lambda_{3}$ and $\theta_{\text{L,out}}
= \lambda_{2} - 3\lambda_{3} + 2\lambda_{4}$ (which are denoted as
$\phi_{1}$ and $\phi_{2}$ by \citealt{Mills2016}) librate about the
observed values, i.e., $\theta_{\text{L,in}} \sim 180\degr$ and
$\theta_{\text{L,out}} \sim 60\degr$.
Here $\lambda_i$ is the mean longitude of planet $i$.
The disk parameters $\Sigma_0$ and $h_0$, the values of $\alpha$ and
$\beta$ surveyed, and the eccentricity damping applied during the
simulations are the same as those used in Section \ref{sec:Two Planets}.
Each simulation is integrated for one migration timescale of the
outermost planet or $1\,$Myr, whichever is smaller.

\begin{figure}
\centering
\includegraphics[width=0.8\linewidth]{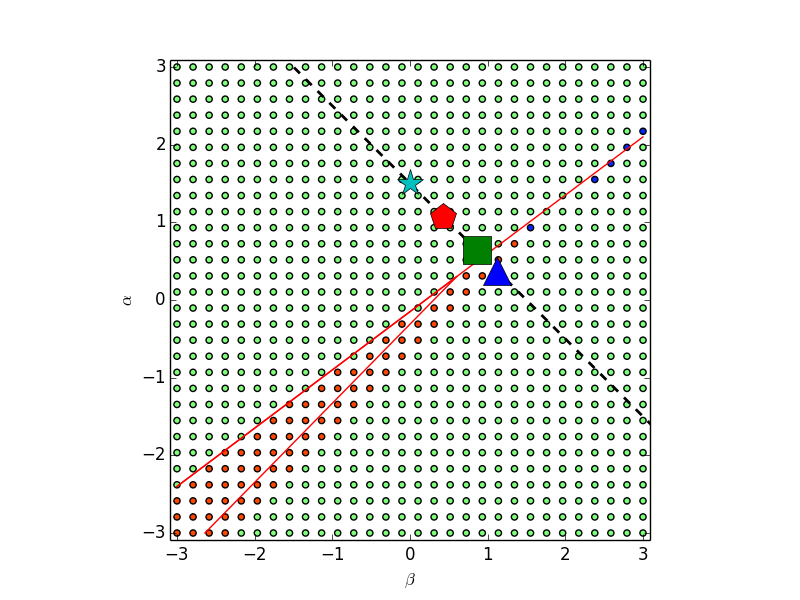}  
\caption{Same as Figure \ref{fig:2_pl_sus} but for the 8:6:4:3
  resonant chain of Kepler-223.
}
\label{fig:223_sus}
\end{figure}

The simulation results are shown in Figure \ref{fig:223_sus}.
As in Figure \ref{fig:2_pl_sus}, the red and blue dots represent the
systems where the resonant chain is maintained by convergent outward
and inward migration, respectively, and the green dots represent the
systems where the resonant chain is broken.
For all the green dots, the period ratios evolve significantly from
the original period ratios, and there are no ambiguous situations
where the period ratios remain close to the original ones but the
resonance angles are not librating.
The region bounded by the two red solid lines is the convergent
outward migration zone found analytically and shown in Figure
\ref{fig:223_analy}(d).
All of the simulations within this analytically derived region are
indeed able to maintain the resonant chain by convergent outward
migration.
However, there are also some systems outside the analytic region with
a sustained resonant chain.
There are some red dots just below the line defined by the constraints
in Equation (\ref{eqs:co_mig_crit_out}), and there are five blue dots
near the extension of the red line representing the boundary between
inward and outward migration given by Equation
(\ref{eqs:disk_constraint}).

\begin{figure}
  \centering
\gridline{\fig{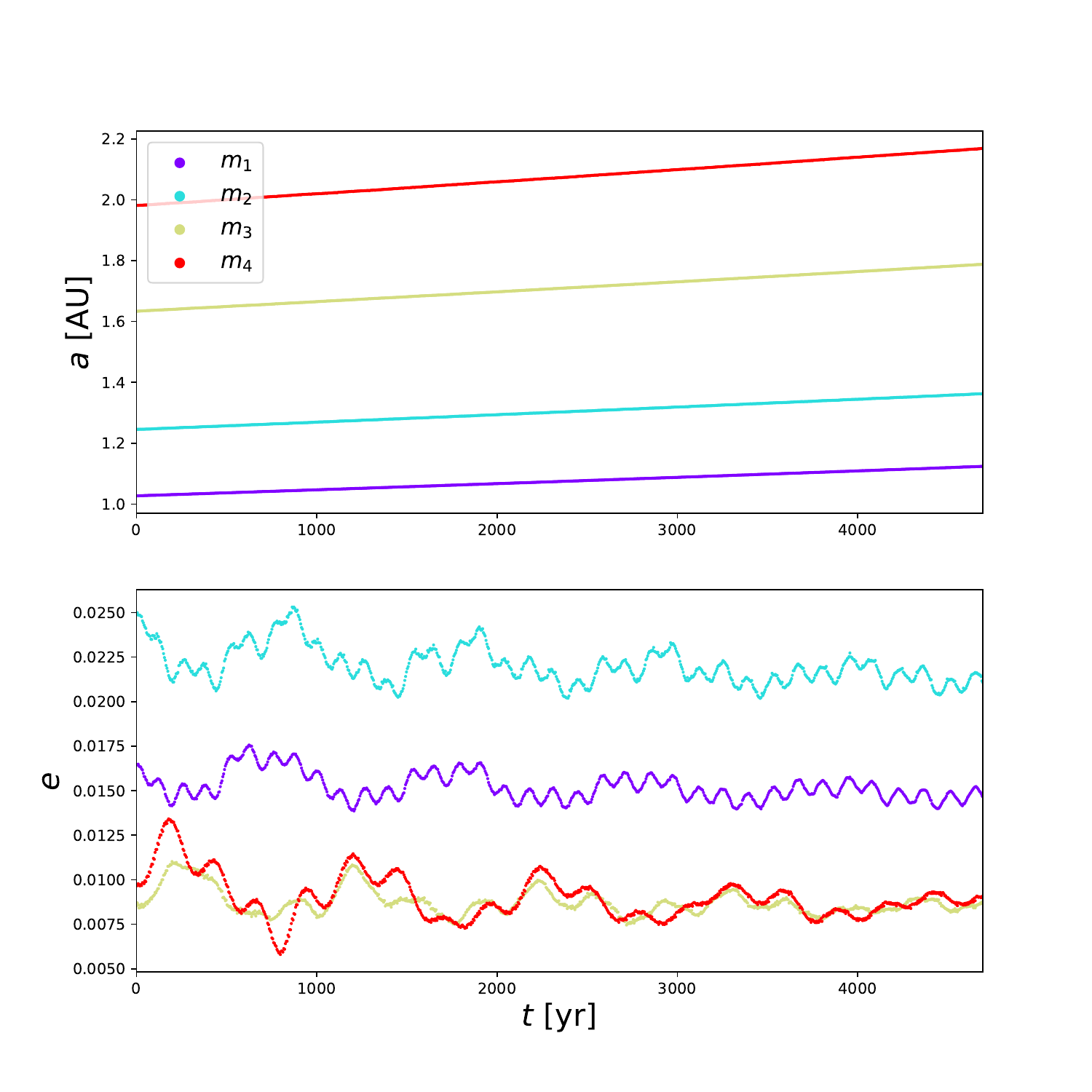}{0.46\textwidth}{}
          \fig{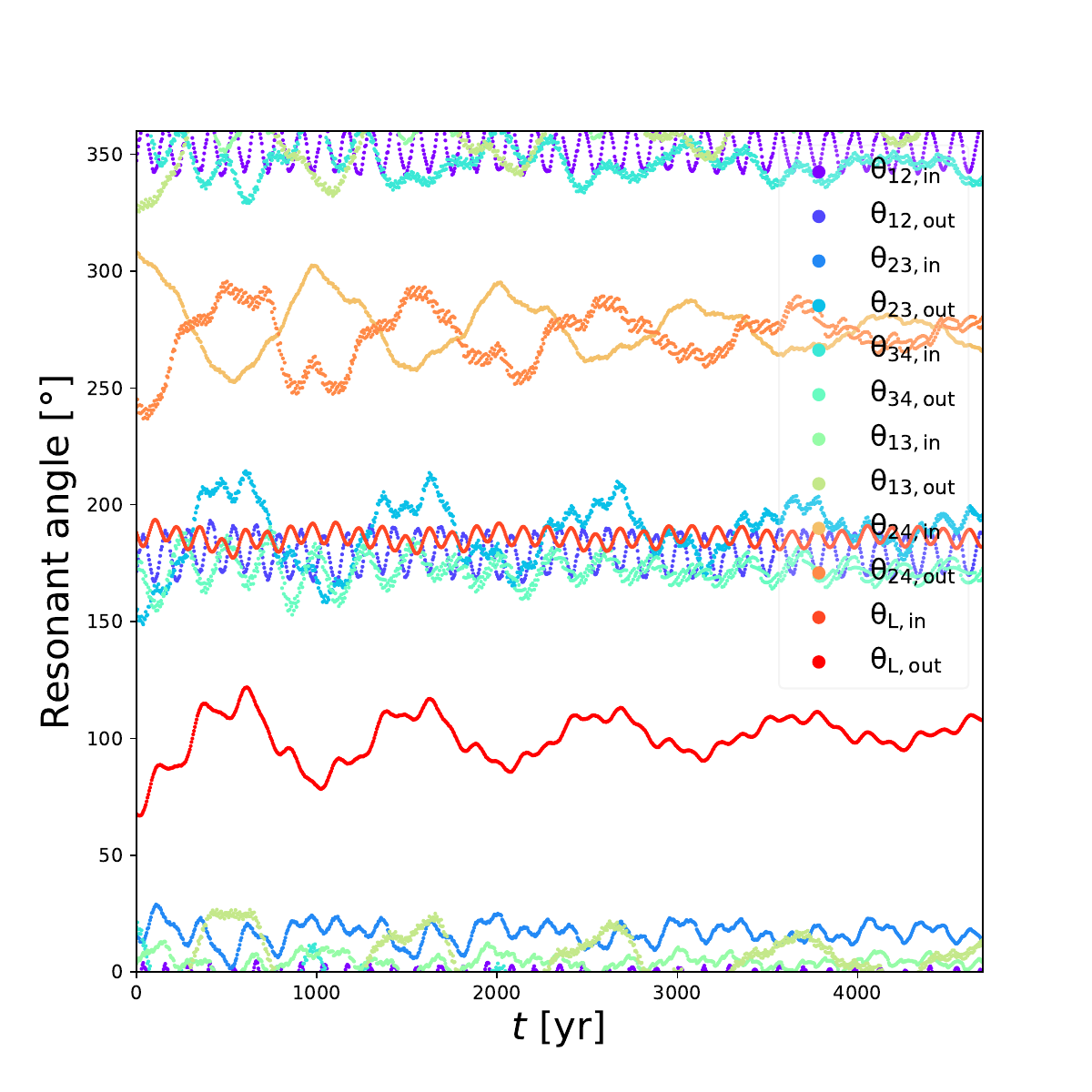}{0.46\textwidth}{}
}
\caption{Time evolution of semimajor axes (upper left panel),
  eccentricities (lower left panel), and two-body MMR and three-body
  Laplace angles (right panel) in a disk migration simulation of the
  Kepler-223 system with $\alpha = -2.586$ and $\beta = -1.966$.
  This $(\alpha,\beta)$ combination is just below the analytic
  convergent outward migration zone, but the resonant chain is
  maintained to the end of the simulation.}
\label{fig:223_unex2}
\end{figure}

Figure \ref{fig:223_unex2} shows the simulation with $\alpha = -2.586$
and $\beta = -1.966$, as an example of the simulations that are just
below the analytic region but still able to maintain the resonant
chain by outward convergent migration.
The evolution of the orbital semimajor axes, eccentricities, and
resonance angles is shown in the upper left, lower left, and right
panels, respectively.
In addition to the three-body Laplace angles $\theta_{\text{L,in}}$
and $\theta_{\text{L,out}}$, we also show the two-body MMR angles
(with $\theta_{\text{12,in}} = 3\lambda_1 - 4\lambda_2 + \varpi_1$ and
$\theta_{\text{12,out}} = 3\lambda_1 - 4\lambda_2 + \varpi_2$, where
$\varpi_i$ is the longitude of periapse of planet $i$, for planets 1
and 2, etc.).
As the four planets migrate outward by $\sim 10\%$, the eccentricities
are rather stable, with slightly decreasing amplitudes of variation
toward the end of the integration, and
all resonance angles remain in libration with no obvious change in the
libration centers.
It is not yet clear why this and other cases just below the analytic
region are able to maintain the resonant chain during outward
migration.
One possibility is that the nonadjacent planets are also in
first-order MMR in this system (both the b-d and c-e pairs are in 2:1
MMR), but further investigation is needed.

\begin{figure}
\centering
\gridline{\fig{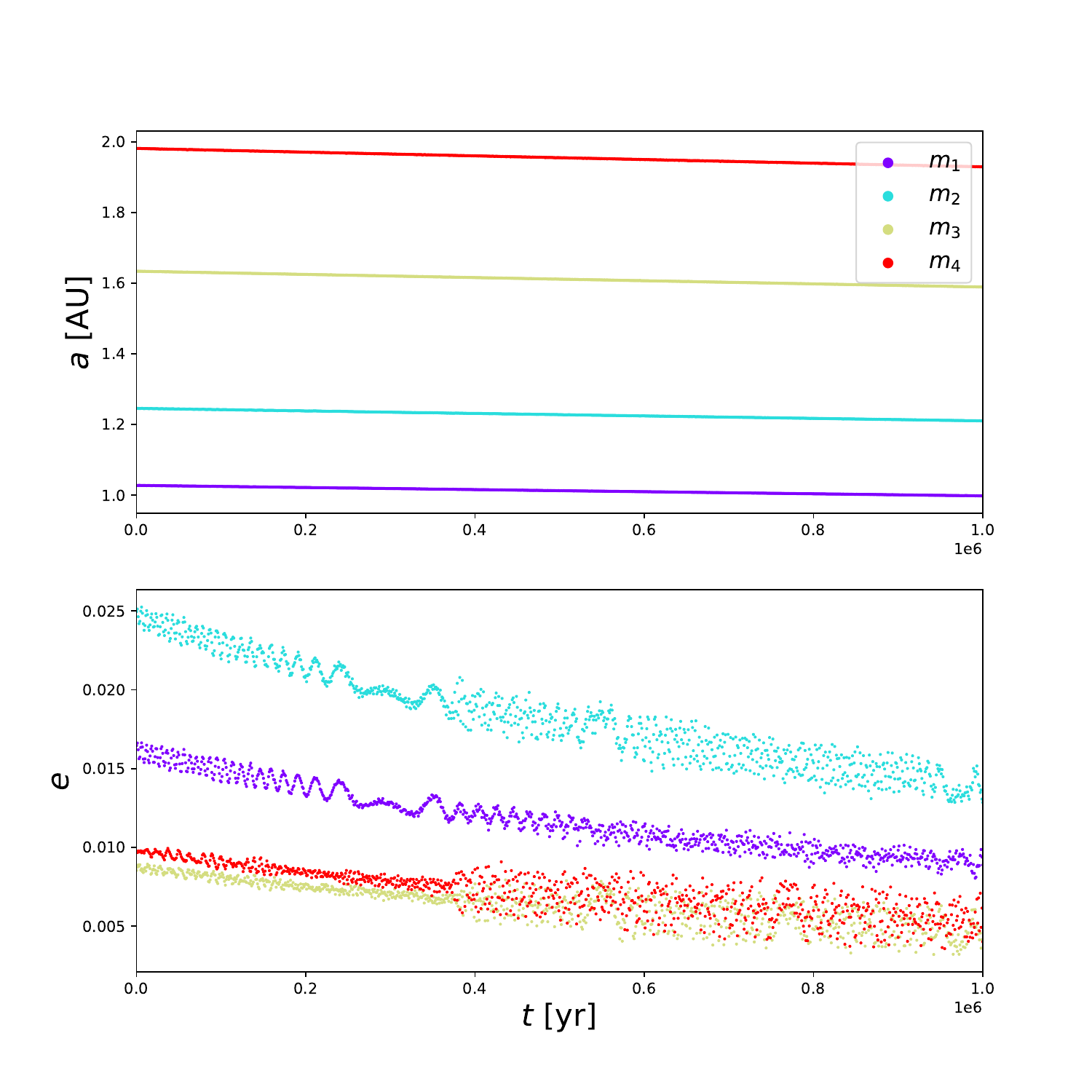}{0.46\textwidth}{}
          \fig{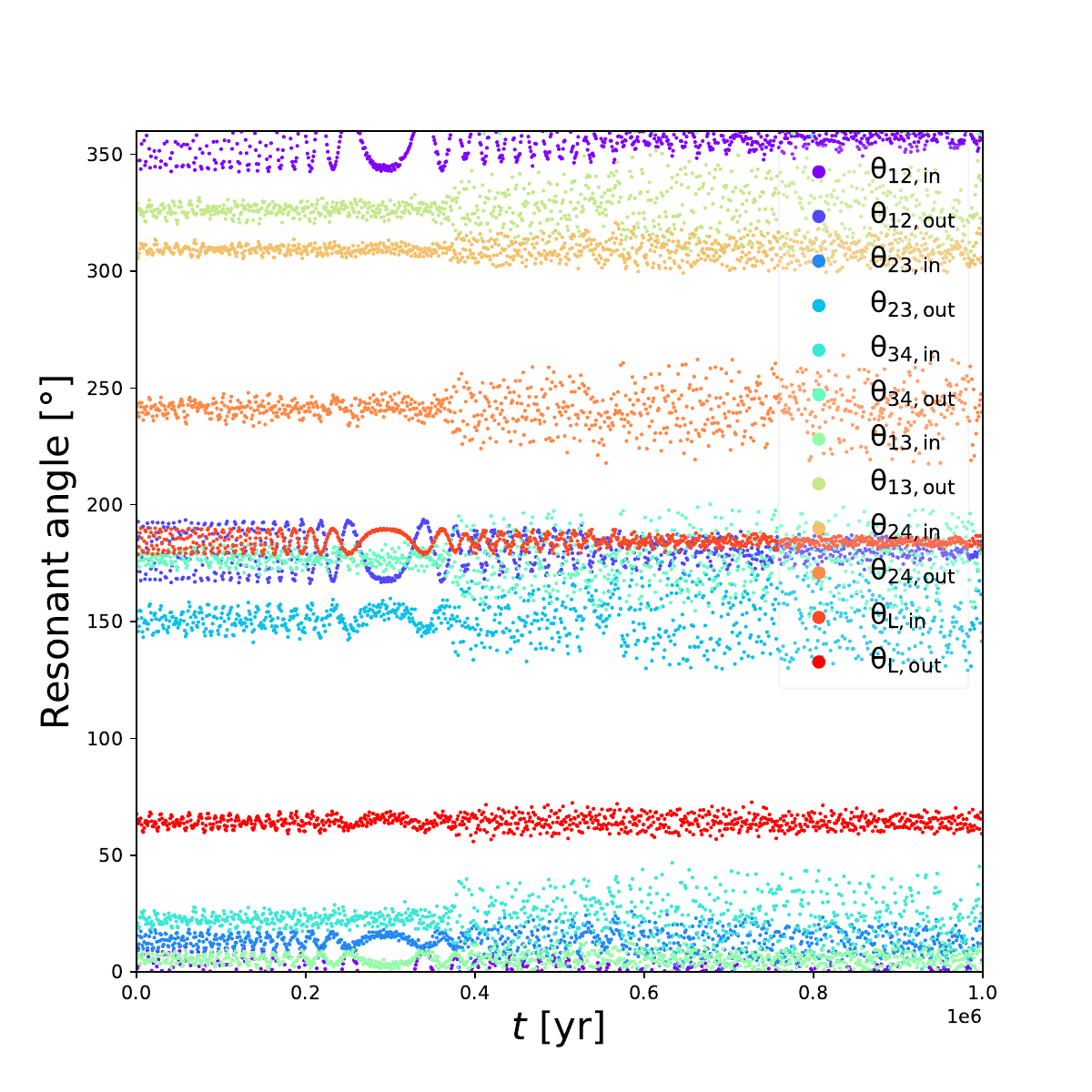}{0.46\textwidth}{}
}
\caption{Same as Figure \ref{fig:223_unex2} but for a simulation with
  $\alpha = 0.931$ and $\beta = 1.552$.
  This $(\alpha, \beta)$ combination is near the line of no migration,
  and the resonant chain is not broken by the very slow inward
  migration after $1\,$Myr.}
\label{fig:223_unex1}
\end{figure}

Figure \ref{fig:223_unex1} shows the evolution of the semimajor
axes, eccentricities, and resonance angles of the simulation with
$\alpha = 0.931$ and $\beta = 1.552$.
This is one of the blue dots near the red line representing the
boundary between inward and outward migration, i.e., the line of no
migration.
The eccentricities decrease over the $1\,$Myr integration, and
some of the resonance angles show an increase in libration amplitude
around $0.4\,$Myr.
But the migration is indeed very slow, so the differential migration
is not significant enough to break the resonant chain.

The numerical results show that the analytic criterion for the
resonant chain in Kepler-223 is not as accurate as in the case of two
planets in MMR.
Nevertheless, the region in $(\alpha, \beta)$ where the resonant chain
is maintained during outward migration is only slightly larger, and
the cases of sustained resonant chains during inward migration are
special cases near the line of no migration with very slow migration.
Therefore, the analytic criterion is still a good indicator of the
types of disk that can maintain a resonant chain during migration.

If the resonant chain in Kepler-223 migrated any significant amount
after assembly, the migration should be inward because the observed
planets are close to the star (with $P_b = 7.38\,$days).
But our analysis shows that there is no region in $(\alpha, \beta)$
where the resonant chain is maintained by inward migration.
Even if we allow for outward migration, the steady-state, constant
$\alpha_\nu$-viscosity disk models (dashed line in Figure
\ref{fig:223_sus}) are outside the region where the resonant chain can
be maintained by outward migration.

\subsection{Application to Other Resonant Chain Systems} \label{sec:Application to Other Resonant Chain Systems}

Now we apply the analytic criterion to several other resonant chain
systems.
For convenience, we label the planets in a resonant chain of $N$
planets in numerical order from $1$ for the innermost planet to $N$
for the outermost planet.

\begin{figure}
  \centering
  \gridline{\fig{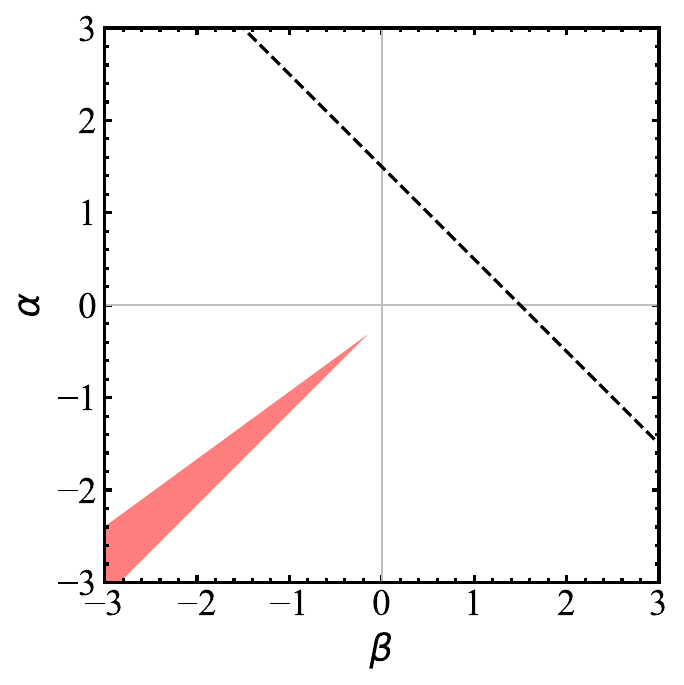}{0.4\textwidth}{(a) Kepler-60}
          \fig{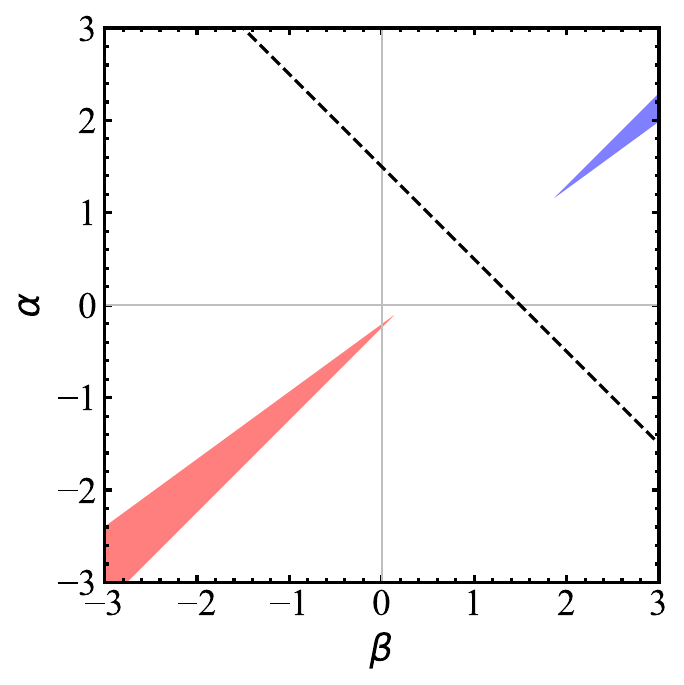}{0.4\textwidth}{(b) Kepler-80}
  }
  \gridline{\fig{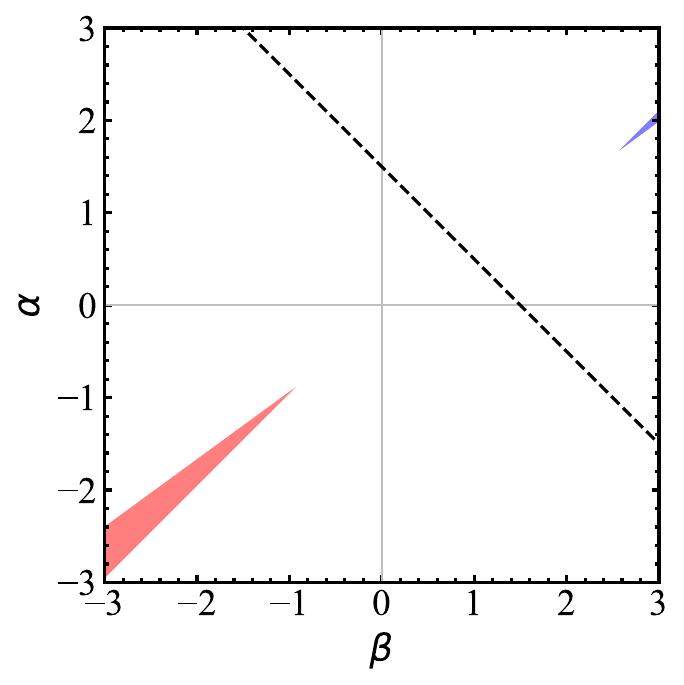}{0.4\textwidth}{(c) TOI-178}
          \fig{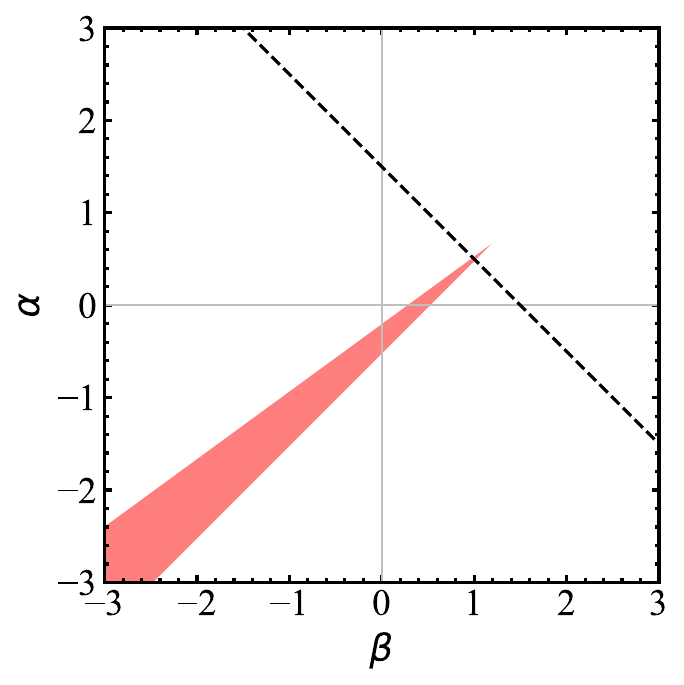}{0.4\textwidth}{(d) TRAPPIST-1}
          }
\caption{Analytic convergent migration zones in $(\alpha, \beta)$
  similar to Figure \ref{fig:223_analy}(d) but for (a) Kepler-60, (b)
  Kepler-80, (c) TOI-178, and (d) TRAPPIST-1.}
\label{fig:Kepler_analy}
\end{figure}

For Kepler-60, the resonant chain consists of three planets in 5:4:3
resonance \citep{Steffen2013}.
We adopt planetary masses of $m_1 = 4.1 M_{\oplus}$, $m_2 =
4.8 M_{\oplus}$, and $m_3 = 3.8 M_{\oplus}$ \citep{Godziewski2016}.
Stellar mass is not needed to determine the locations of the
convergent migration regions.
Figure \ref{fig:Kepler_analy}(a) shows the analytic result, which
is the overlapping region from the convergent migration zones
between pair 1-2 and planet 3 and between planet 1 and pair 2-3.
There is no inward convergent migration zone that can maintain the
resonant chain, as in Kepler-223.
The outward convergent migration zone is slightly smaller than that
for Kepler-223 and requires negative values of $\alpha$ and $\beta$.

Kepler-80 has six known planets, with all but the innermost planet f
in a resonant chain.
The outermost planet g cannot be included in our analysis, because its
mass is not known \citep{Shallue2018}.
The middle four planets are in 9:6:4:3 resonance.
We adopt planetary masses of $m_1 = 6.75 M_{\oplus}$, $m_2 =
4.13 M_{\oplus}$, $m_3 = 6.93 M_{\oplus}$, and $m_4 = 6.74 M_{\oplus}$
\citep{MacDonald2016}.
Figure \ref{fig:Kepler_analy}(b) shows that the outward convergent
migration zone is similar to that of Kepler-223 and that there is a
small region of inward convergent migration around $\alpha \sim 2$ and
$\beta \sim 3$ that can maintain the resonant chain.

TOI-178 is also a six-planet system with all but the innermost planet
b in a resonant chain.
The resonance is 18:9:6:4:3.
We adopt planetary masses of $m_1 = 4.77 M_{\oplus}$, $m_2 = 3.01
M_{\oplus}$, $m_3 = 3.86 M_{\oplus}$, $m_4 = 7.72 M_{\oplus}$, and
$m_5 = 3.94 M_{\oplus}$ \citep{Leleu2021}.
Figure \ref{fig:Kepler_analy}(c) shows that both inward and outward
convergent migration zones are smaller than those of Kepler-80.
Since the type I forced migration rate is proportional to planetary
mass, the particularly massive $m_4$ in the middle of the chain makes
it more difficult for the whole chain to maintain convergent
migration.

For the seven-planet system TRAPPIST-1, we adopt planetary masses of
$m_1 = 1.374 M_{\oplus}$, $m_2 = 1.308 M_{\oplus}$, $m_3 = 0.388
M_{\oplus}$, $m_4 = 0.692 M_{\oplus}$, $m_5 = 1.039 1M_{\oplus}$, $m_6
= 1.321 M_{\oplus}$, and $m_7 = 0.326 M_{\oplus}$ \citep{Agol2021}.
Although the mean motions of the inner two planets b and c are nearly
in the ratio 8:5:3 with that of planet d, it is not clear that they
are part of the chain of first-order resonances (9:6:4:3:2) of the
outer five planets \citep{Gillon2017, Luger2017}.
Figure \ref{fig:Kepler_analy}(d) shows the analytic result if we
assume that all seven planets are in the resonant chain.
Due to the innermost planet b ($m_1$) being the most massive planet in
this system, the outward convergent migration zone is the largest for
all the systems considered.
There is no inward convergent migration zone that can maintain the
resonant chain, because the outermost planet h ($m_7$) is
significantly less massive than the rest of the planets (except
$m_3$), which makes it difficult for $m_7$ to catch up with the rest
of the planets in inward migration.
If we assume that planets b and c are not in the resonant chain, we
find that no region in $(\alpha, \beta)$ space can maintain the
resonant chain in either inward or outward convergent migration,
because both the innermost planet d ($m_3$) and the outermost planet h
($m_7$) of the resonant chain are now significantly less massive than
the rest of the planets.

As shown in Section \ref{sec:Application to Kepler-223}, the analytic
criterion is a good indicator of the types of disk that can maintain a
resonant chain during migration. even though the regions of inward and
outward convergent migration in $(\alpha, \beta)$ may be slighter
larger than the analytic ones.
The analytic outward migration zone is the largest for TRAPPIST-1 if
we assume that all seven planets are in the resonant chain.
But even in this case, it just touches the steady-state, constant
$\alpha_\nu$-viscosity disk models (dashed line in Figure
\ref{fig:Kepler_analy}).
In all cases, the resonant chain can be maintained by inward
convergent migration in, at most, a small region of $(\alpha,
\beta)$.
As in the case of Kepler-223, the planets in the systems discussed in
this section are close to the stars, and the difficulty with inward
migration is particularly problematic.

\section{Effects of Uncertainties in Planetary Masses}
\label{sec:Uncertainties}

All of the resonant chain systems examined above were discovered by
the transit method, and the uncertainties in the planetary masses can
be as much as $\sim 40\%$.
As the type I migration rate of a planet scales linearly with its
mass, the mass uncertainties can significantly affect the estimation
of convergent migration zones in $(\alpha, \beta)$.
In this section, we examine how the convergent migration zones are
changed if we adjust the planetary masses within $1\sigma$.
To enhance the inward convergent migration zone, we lower the mass of
the innermost planet by $1\sigma$ and raise the mass of the
outermost planet by $1\sigma$.
For the other planets in the chain, their masses are adjusted within
$1\sigma$ to archive a relatively smooth mass variation.

\begin{figure}
  \centering
  \gridline{\fig{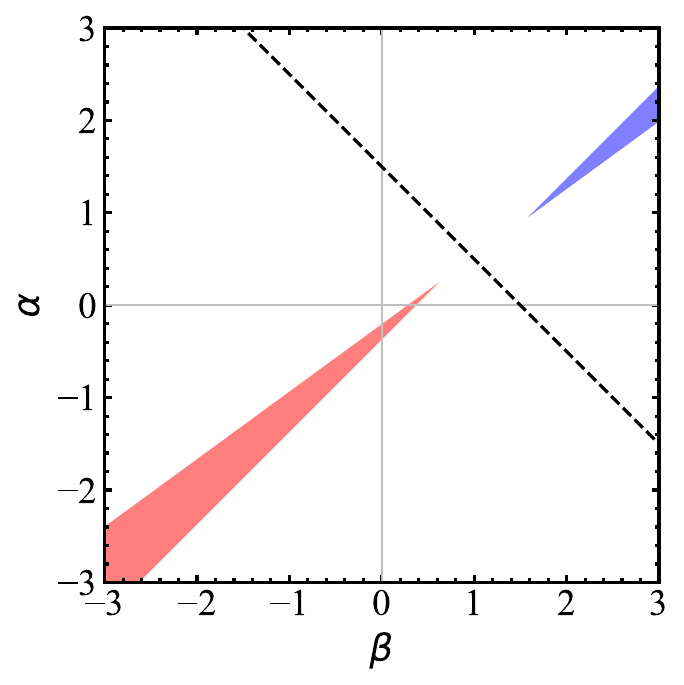}{0.4\textwidth}{(a) Kepler-223}
    \fig{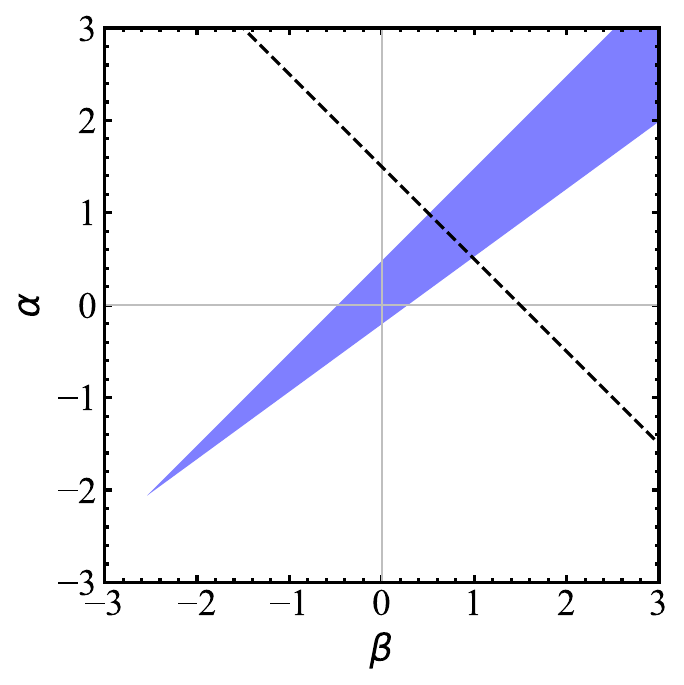}{0.4\textwidth}{(b) Kepler-60}
  }
  \gridline{\fig{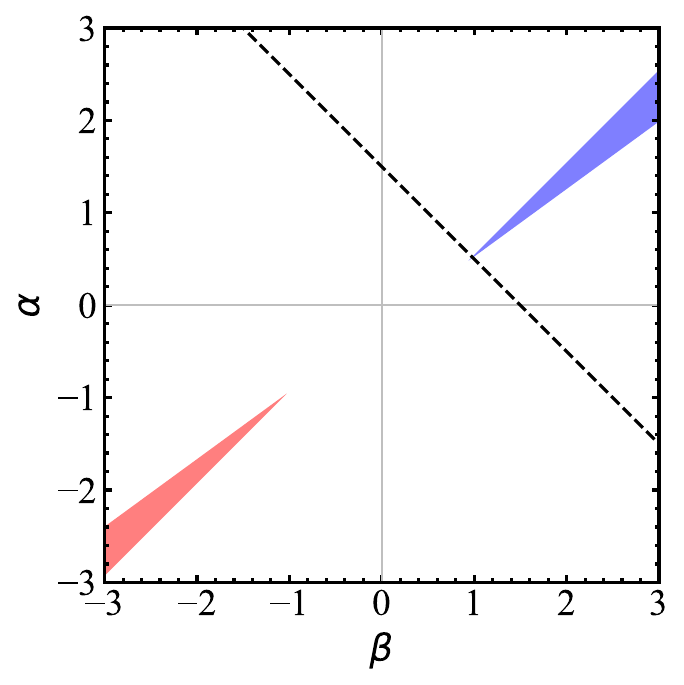}{0.4\textwidth}{(c) Kepler-80}
    \fig{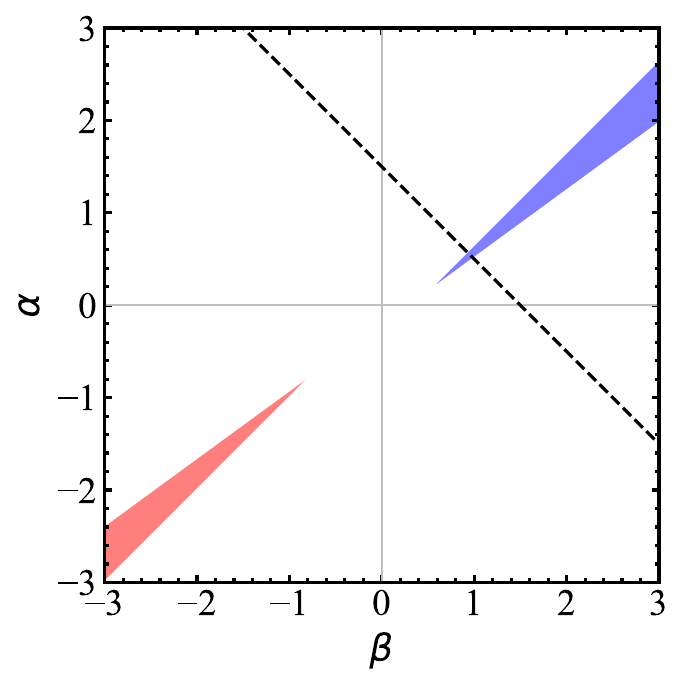}{0.4\textwidth}{(c) TOI-178}
  }
\caption{Analytic convergent migration zones in $(\alpha, \beta)$
  similar to Figures \ref{fig:223_analy}(d) and
  \ref{fig:Kepler_analy} but for planetary masses adjusted within
  $1\sigma$ to enhance the inward convergent migration zone.}
\label{fig:Kepler_analy_newm}
\end{figure}

For Kepler-223, the adjusted masses are $m_1 = 6.3 M_{\oplus}$, $m_2 =
6.8 M_{\oplus}$, $m_3 = 6.7 M_{\oplus}$, and $m_4 = 6.2 M_{\oplus}$.
The analytic convergent migration zones with the adjusted masses are
shown in Figure \ref{fig:Kepler_analy_newm}(a).
Compared to the nominal mass result shown in Figure
\ref{fig:223_analy}(d), the outward convergent migration zone is
almost identical, but there is now a small inward
convergent migration zone.
For Kepler-60, we adjust the masses to $m_1 = 3.4 M_{\oplus}$, $m_2 =
3.8 M_{\oplus}$, and $m_3 = 4.9 M_{\oplus}$.
With $m_3 > m_2 \sim m_1$, the outward convergent migration zone in
Figure \ref{fig:Kepler_analy}(a) is replaced by a large inward
migration zone in Figure \ref{fig:Kepler_analy_newm}(b).
For Kepler-80, the adjusted masses are $m_1 = 6.24 M_{\oplus}$, $m_2 =
4.94 M_{\oplus}$, $m_3 = 6.23 M_{\oplus}$, and $m_4 = 7.97 M_{\oplus}$.
Comparing the results with nominal masses in Figure
\ref{fig:Kepler_analy}(b) and adjusted masses in Figure
\ref{fig:Kepler_analy_newm}(c), the inward migration zone is larger,
while the outward migration zone is smaller.
For TOI-178, we adjust the masses to $m_1 = 4.09 M_{\oplus}$, $m_2
= 3.01 M_{\oplus}$, $m_3 = 3.86M_{\oplus}$, $m_4 = 6.20 M_{\oplus}$,
and $m_5 = 5.25M_{\oplus}$.
Comparing Figure \ref{fig:Kepler_analy}(c) to Figure
\ref{fig:Kepler_analy_newm}(d), the outward convergent migration zone
is almost unchanged, while the inward convergent migration zone is
larger.
Finally, for TRAPPIST-1, the uncertainties in the masses reported by
\cite{Agol2021} are sufficiently small ($6\%$ or less) that the
results are almost the same for any variation of the masses within
$\pm 1\sigma$: the analytic convergent migration zone is almost the
same as in Figure \ref{fig:Kepler_analy}(d) if all seven planets are
in the resonant chain, and no region in $(\alpha, \beta)$ space can
maintain the resonant chain in either inward or outward convergent
migration if planets b and c are not in the resonant chain.

These examples show that even if we take into account the
observational uncertainties in the planetary masses to optimize the
size of the $(\alpha, \beta)$ region where the resonant chain can be
sustained by inward convergent migration, there is very little or no
overlap between the inward convergent migration zone (shaded in blue
in Figure \ref{fig:Kepler_analy_newm}) and the steady-state, constant
$\alpha_\nu$-viscosity disk models (dashed line with $\alpha + \beta =
3/2$ in Figure \ref{fig:Kepler_analy_newm}) in most cases.
In fact, within the observational uncertainties, there is no inward
convergent migration zone for TRAPPIST-1 due to the small mass of the
outermost planet h.
The only exception is Kepler-60, where, within uncertainties, it is
possible to have a substantial overlap between the inward convergent
migration zone and the steady-state, constant $\alpha_\nu$-viscosity
disk models.
Therefore, a more accurate determination of planetary masses in the
resonant chain systems would further constrain their origin and
migration in a disk.

\section{Formation of Resonant Chain near the Inner Disk Edge}\label{sec:Scaling}

We have demonstrated that the observed resonant chains are difficult
to maintain in reasonable disk models if they migrate any significant
amount after assembly.
In this section, we show that a relationship between the orbital
semimajor axis of the innermost planet and the stellar mass of the
observed resonant chains supports the idea that they have a common origin
in a stalling and assembly mechanism near the inner edge of the
protoplanetary disk.

In a simple model of a disk with an inner edge, the surface density
is zero at the inner edge at $a_{\text{tr}}$, rises to a maximum at
some distance from the inner edge, and then decreases away from the
star.
This means that the power-law index $\alpha = - d\ln \Sigma/d\ln a$
would change from highly negative near the inner edge to zero at the
surface density maximum and then positive far from the inner edge.
For any reasonable variation of the temperature power-law index $\beta
= - d\ln T/d\ln a$, there is a zero-torque location at some distance
from the inner edge, where $\alpha$ and $\beta$ satisfy Equation
(\ref{eqs:disk_constraint}), the type I migration torque is zero, and
a planet would stop migrating.
At distances closer to (farther from) the star, the torque would be
positive (negative), and a planet would migrate outward (inward).
As the innermost planet migrates inward, it would be trapped at the
zero-torque location.
The outer planets would continue their inward migration and would be
naturally captured into resonance with the inner planets,
regardless of the variation in planetary masses.
As the resonant chain is assembled, the innermost planet would move
slightly inward from the zero-torque location, so that its outward
migration would balance the inward migration of the outer planets.

\begin{figure}
\centering
\includegraphics[width=0.75\linewidth]{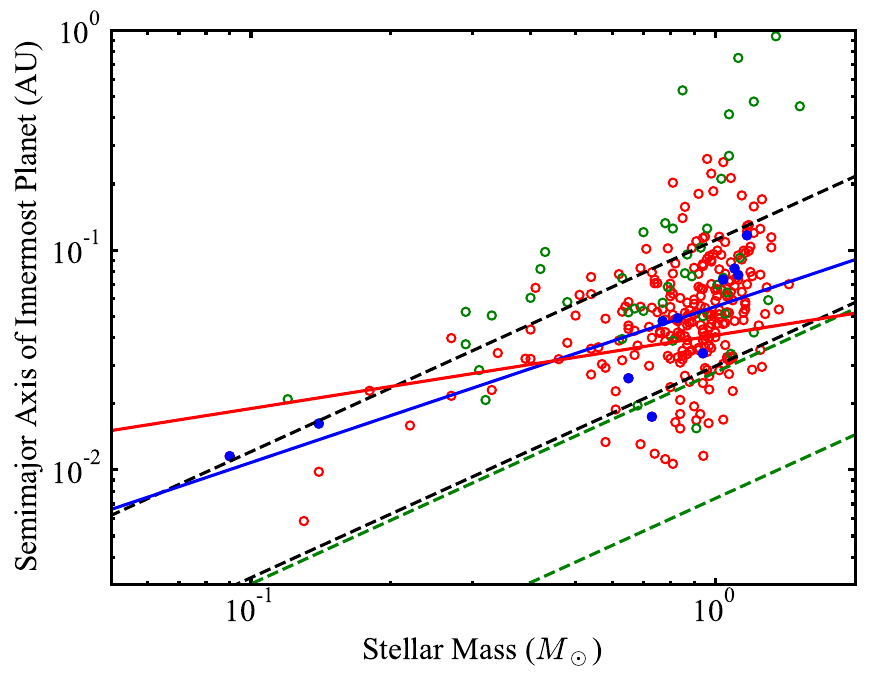}  
\caption{Orbital semimajor axis of innermost planet versus stellar
  mass for detected planetary systems with three or more planets, with
  red and green open circles for the planets discovered by transit and
  radial velocity, respectively, and blue filled circles for 11
  confirmed and suspected resonant chain systems (see text for the
  list).
  The green dashed lines are the upper and lower estimates of the
  magnetic truncated radius $a_{\text{tr}}$ of the protoplanetary disk
  (Equation (\ref{eqs:atrun})), and the black dashed lines are the
  upper and lower estimates of $4 a_{\text{tr}}$.
  The solid red line is $a_{\text{tr}}$ derived by \cite{Batygin2023}.
  The solid blue line is the best-fit straight line to the blue filled
  circles.
  Data obtained from NASA Exoplanet Archive.}
\label{fig:d_vs_m}
\end{figure}

This idea of forming the resonant chain near the inner disk edge is
also supported by the positions of the innermost planets in the
observed resonant chain systems.
In Figure \ref{fig:d_vs_m}, we show the orbital semimajor axis of the
innermost planet versus the stellar mass for two populations of planets.
The blue filled circles are the innermost planets in $11$ confirmed and
suspected resonant chain systems.
In addition to Kepler-60, Kepler-80, Kepler-223, TOI-178, and
TRAPPIST-1,
we also include HD~40307 \citep{Diaz2016}, Kepler-79
\citep{Jontof2014, Yoffe2021}, K2-32 \citep{Heller2019, LilloBox2020},
K2-138 \citep{Christiansen2018, Lopez2019}, V1298 Tau
\citep{David2019b, SuarezMascareno2022}, and YZ Ceti
\citep{Stock2020}.
Some of these are suspected but not confirmed resonant chains, and
HD~40307 is suspected to be in a resonant chain in the past
\citep{Papaloizou2010}.
GJ~876 and HR~8799 are two resonant chain systems not included in this
sample because of the presence of giant planets that undergo type II
instead of type I migration.
The open circles are the innermost planets in other systems with three or
more planets, with red and green open circles for the planets
discovered by transit and radial velocity, respectively.

For comparison, if the inner cavity of the protoplanetary disk is
created by the magnetic fields of the protostar, the truncation radius
is given by the balance between the stellar magnetic stress and the
disk Reynolds stress \citep{Starczewski2007, Chang2010}:
\begin{equation} \label{eqs:atrun}
\frac{a_{\text{tr}}}{R_{\odot}} = 4.29 \eta \left[
  \frac{(B_{*}/1000\,\text{G})^4 (R_{*}/R_{\odot})^{12}}
       {(M_{*}/M_{\odot}) ({\dot M}_{\text{disk}}/10^{-8}
         M_{\odot}\,\text{yr}^{-1})^2}
  \right]^{1/7} ,
\end{equation}
where $\eta$ is a dimensionless factor of order unity,
$B_{*}$ is the strength of the magnetic field at the stellar surface,
$R_{*}$ and $M_{*}$ are the stellar radius and stellar mass, and
${\dot M}_{\text{disk}}$ is the disk's accretion rate.
If we adopt $0.5 < \eta < 1$ for a star with an aligned dipole field,
$500\,\text{G} < B_{*} < 1500\,\text{G}$, the stellar mass-radius
relation $R_\ast/R_\odot = 1.06 (M_\ast/M_\odot)^{0.945}$
\citep{Demircan1991}, and ${\dot M}_{\text{disk}} = 10^{-8}
(M/M_{\odot})^{1.8} M_{\odot}\,\text{yr}^{-1}$
\citep{Muzerolle2003, Natta2006, Manara2016}, we have $a_{\text{tr}}
\propto M_{\ast}^{0.963}$ and the estimated range of $a_{\text{tr}}$
is shown as the upper and lower green dashed lines in Figure
\ref{fig:d_vs_m}.
Since the zero-torque location is expected to be at some distance from
the inner edge, we also show the estimated range of $4 a_{\text{tr}}$
as the black dashed lines.

As seen in Figure \ref{fig:d_vs_m}, the orbital semimajor axis of the
innermost planet of the resonant chain systems (blue filled circles)
shows a scaling relationship with the stellar mass that is roughly
bounded by the two black dashed lines, which supports the idea that
the resonant chains are assembled by halting the migration of the
innermost planet at the zero-torque location near the inner disk edge.
For the nonresonant chain multiplanet systems (open circles), many
fall within or just outside the region bounded by the two black dashed
lines, but some are significantly above the upper black dashed lines.
This would be consistent with the former corresponding to systems where
the innermost planet is stopped at the zero-torque location but the
other planets in the system do not converge on it sufficiently to
form a resonant chain and the latter corresponding to systems where
the innermost planet stops migrating (due to, e.g., disk dispersal)
before it reaches the zero-torque location.

There are significant uncertainties on how $B_\ast$, $R_\ast$, and
${\dot M}_{\text{disk}}$ scale with $M_\ast$.
With different assumptions ($B_\ast \propto M_\ast^{1/3}$, $R_\ast
\propto M_\ast^{1/3}$, and ${\dot M}_{\text{disk}} \propto M_\ast$),
\cite{Batygin2023} have recently argued that $a_{\text{tr}} \propto
M_\ast^{1/3}$, which corresponds to an orbital period of $\sim 3.0$
days independent of $M_\ast$ (red solid line in Figure
\ref{fig:d_vs_m}).
There is sufficient scatter in the resonant chain data shown in Figure
\ref{fig:d_vs_m} that they could be consistent with either
$a_{\text{tr}} \propto M_\ast^{0.963}$ or $M_\ast^{1/3}$.
If we fit a straight line to the data, we find that $a \propto
M_\ast^\mu$ where $\mu = 0.71 \pm 0.17$ (solid blue line in Figure
\ref{fig:d_vs_m}).
In the future, as more and more resonant chain systems are discovered,
if they continue to fall on the currently observed trend of orbital
semimajor axis of the innermost planet versus stellar mass ($a \propto
M_\ast^\mu$ where $\mu \approx 0.333$--$0.963$), the case for a common
origin near the inner disk edge would be strengthened.


\section{Conclusions}\label{sec:Conclusions}

Convergent migration is a necessary criterion to maintain planets in a
resonant chain, if the resonant chain is formed before the dispersal
of the protoplanetary gas disk and the planets continue to migrate due
to planet-disk interactions.
Therefore, we have investigated what kind of protoplanetary disk would
allow such convergent migration. 
The planets are assumed to undergo type I migration in an adiabatic
disk with power-law indices $\alpha = - d\ln \Sigma/d\ln a$ and $\beta
= - d\ln T/d\ln a$.
We have developed an analytic criterion to determine the convergent
migration zone in the $(\alpha,\beta)$ parameter space.
For a pair of planets in MMR, the convergent migration zone is bounded
by two lines: the disk constraint (which is the boundary between
inward and outward migration) and the planetary mass constraint (which
is the boundary between faster inner planet migration and faster outer
planet migration).
The criterion was generalized to a resonant chain by requiring that
any part of the resonant chain should be convergently migrating toward
the remaining part.
We have verified the analytic criterion with numerical simulations of
a pair of $1 M_{\oplus}$ planets in 3:2 MMR and the four-planet
resonant chain in the Kepler-223 system.
The analytic criterion was then applied to the Kepler-60, Kepler-80,
TOI-178, and TRAPPIST-1 systems.
In all cases, outward convergent migration requires rather extreme
(mostly negative) values of $(\alpha, \beta)$, and there is little or
no inward convergent migration zone.
Even when we adjust the planetary masses within the uncertainties to
maximize the inward convergent migration zone,
there is little or no overlap between the inward convergent migration
zone and common disk models (such as the steady-state, constant
$\alpha_\nu$ disks) for all but one of the observed systems.

For TRAPPIST-1, the masses of the planets are well determined, and the
biggest uncertainty in the existence and sizes of the convergent
migration zone is whether the inner two planets b and c are (or were)
in the resonant chain.
For other resonant chains (in particular, Kepler-60), the
uncertainties in the existence and sizes of the inward and outward
convergent migration zones in $(\alpha, \beta)$ space can be improved
by reducing the uncertainties in the planetary masses.
In our analysis, we have assumed unsaturated nonlinear type I
migration torque.
The formulae developed by \cite{Paardekooper2011} can be used if
the torques on the planets in a particular resonant chain system are
significantly affected by the effects of viscous and thermal
diffusion.
However, the formulae depend on the viscous saturation parameter
$p_\nu$ and the thermal saturation parameter $p_\chi$, and it will not
be possible to derive a simple analytic criterion for convergent
migration.
In fact, the criterion could vary as a function of position due to the
functional form of $p_\chi$ (see Section \ref{sec:Disk Model}).

The innermost planets in the observed resonant chain systems are close
to the stars, and an alternative scenario is that the resonant chain
is formed and maintained by the inward migration of the outer planets
toward the innermost planet after the migration of the latter is
stalled near the inner disk edge.
We have found support for this idea from a relationship between the
orbital semimajor axis of the innermost planet and the stellar mass of
the observed resonant chain systems.
If resonant chains discovered in the future continue to fall on this
relationship, the case for a common origin near the inner disk edge
will be strengthened.

\acknowledgments
This work was supported by a postgraduate studentship at the
University of Hong Kong (K.H.W.) and Hong Kong RGC grant 17305618
(K.H.W. and M.H.L.).

\bibliography{bibliography}  

\end{document}